\documentclass[11pt,a4paper]{article}
\usepackage{amsmath,amssymb}
\usepackage{cite}
\usepackage{graphicx}
\usepackage{hyperref}
\hypersetup{
	colorlinks=false,
	pdfborder={0 0 0}
}
\def\Rnum#1{\uppercase\expandafter{\romannumeral #1}}
\begin{document}
\begin{flushright}
	KANAZAWA-20-04\\
	September, 2020
\end{flushright}
\vspace*{1cm}

\begin{center}
{\Large\textbf{Inflation and DM phenomenology in a scotogenic model extended with a real singlet scalar}}
\vspace*{1cm}

{\Large Tsuyoshi Hashimoto}\footnote[1]{t\_hashimoto@hep.s.kanazawa-u.ac.jp}
{\Large and Daijiro Suematsu}\footnote[2]{suematsu@hep.s.kanazawa-u.ac.jp}
\vspace*{0.5cm}\\

\textit{Institute for Theoretical Physics, Kanazawa University,
Kanazawa 920-1192, Japan}
\end{center}
\vspace*{1.5cm}

\noindent
{\Large \textbf{Abstract}}\\
We study an extension of the scotogenic model with a real singlet scalar. It gives an origin of the mass of right-handed neutrinos and plays a role of inflaton through a nonminimal coupling with Ricci scalar.  While an inert doublet scalar is an indispensable ingredient for neutrino mass generation in the model, it is also a promising dark matter (DM) candidate. Introduction of the singlet scalar could affect its nature of DM if mass of the singlet scalar is in a resonance region. We focus our study on such a case where inflaton mass is expected to be in a TeV range and reheating temperature is less than $10^9$ GeV. Thus the model requires low scale leptogenesis. After examining several effects brought about by the singlet scalar for the DM sector, we discuss DM phenomenology such as high energy neutrinos and monochromatic gammas caused by its annihilation.

\newpage
\section{Introduction}
%
The standard model (SM) can describe well the nature up to the weak scale. On the other hand, now we know several experimental and observational results which are not explained within it. They are the existence of small neutrino mass \cite{nexp,t13}, dark matter (DM) \cite{dm,dmobs} and baryon number asymmetry in the Universe \cite{baryon}. These require some extension of the SM. As such an extension at TeV scales, we have a simple model called the scotogenic model \cite{ma}, which connects the neutrino mass generation and the existence of DM. In this model, the SM is extended by an inert doublet scalar and right-handed neutrinos. If we assume that these new fields are assigned odd parity under the $Z_2$ symmetry and all the SM contents have even parity, the neutrino mass is generated at one-loop level and the lightest neutral $Z_2$ odd particle can behave as DM. In the original model and its several extensions \cite{fcnc,flavor,sty,u1,ma_type,nonth,ks,bks},\footnote{In the scotogenic model or some modified ones \cite{dma}, possibility of freeze-in DM has been discussed. We focus our study on the parameter region which realizes freeze-out DM here.} various phenomenological issues including the explanation of baryon number asymmetry through leptogenesis \cite{leptg} have been extensively studied.

In this paper, we consider an extension of the model from a viewpoint of cosmological inflation. Cosmic microwave background (CMB) observations suggest that the exponential expansion of the Universe should occur before the ordinary big bang of the Universe \cite{planck,planck18}. On the other hand, analyses of the CMB data seem to have already ruled out a lot of inflation models proposed by now. Higgs inflation is a well-known example which is still alive \cite{higgsinf}. It uses a feature such that Higgs potential becomes flat enough for large field regions if the Higgs scalar has a large nonminimal coupling with Ricci scalar.  We apply this idea to a real singlet scalar which is introduced to the model in order to explain an origin of the mass of right-handed neutrinos.  If the singlet scalar is supposed to have a substantial nonminimal coupling with Ricci scalar, it could work as inflaton. Such a coupling of a real singlet scalar has been studied as $s$-inflation in a different context \cite{sinfl,ext-s}. There, the unitarity problem which appears in the Higgs inflation and many other models \cite{unitarity1,unitarity2} is suggested to be escapable. An interesting point in this extended model is that the inflaton could affect DM phenomenology.

Our Universe is considered to be filled with unknown neutral particles called DM on the basis of several observational results \cite{dmobs}, that is, rotation curves of galaxies, fluctuation of the CMB, bullet clusters and so on. Since the SM has no candidate for it, DM is one of the crucial signatures for physics beyond the SM. DM has been studied through direct search experiments, indirect search experiments, and collider experiments. However, unfortunately we have not found its signature through them still now. Now, direct search experiments put severe constraints on a cross section between a nucleon and DM \cite{crbound}. If we suppose the present DM abundance in the Universe to be explained as thermal relics after their decoupling, the annihilation cross section tends to be larger than the bound obtained by the direct search experiments. Since the DM-nucleon scattering can be directly related to DM-DM annihilation processes in a lot of DM models, they face severe constraints. On the other hand, if the model has no direct relation between interactions which induce the DM nucleon scattering and the DM-DM annihilation, it could open a new possibility for DM phenomenology.

We study inert doublet DM in this extended model as such a candidate. It is known that the relic abundance is determined by coannihilation among components of an inert doublet scalar in the DM mass region such as $\gtrsim 600$~GeV \cite{highmass,idm}. On the other hand, the nucleon-DM scattering is caused only through Higgs exchange, which gives a subdominant contribution in the DM coannihilation. This feature could weaken the above-mentioned tension and keep it as a promising candidate for DM. This DM candidate has other noticeable features also. First, the self-interaction of inert doublet DM could be large enough by the existence of the real singlet scalar if a certain condition is satisfied. In such a case, the DM-DM scattering is enhanced so that the DM capture rate through the self-scattering in the Sun might be affected. Second, the mass of a real part and an imaginary part of the neutral component is favored to be degenerate from a viewpoint of the small neutrino mass generation, and then inelastic scattering could be caused easily for this DM. These are expected to give a crucial influence on DM phenomenology in the model. Taking account of them, we reconsider inert doublet DM physics focusing on high energy neutrinos and monochromatic gammas caused through the DM annihilation \cite{capt0}.

The remaining parts of this paper are organized as follows. In Sec. \Rnum{2}, we briefly explain the extended model studied in this paper. After a possible inflation scenario is discussed, leptogenesis is examined under low reheating temperature realized in the inflation scenario. We reexamine the inert doublet DM in the model and discuss its several features quantitatively. In Sec. \Rnum{3}, an allowed parameter space is also examined by combining the constraints from the DM relic abundance and the DM direct search. Taking account of the DM capture rate by the Sun which could be modified by the nature of the present DM, we estimate expected high energy neutrinos from the Sun. We also study high energy gammas from Galactic Center and dwarf spheroidal galaxies. Section \Rnum{4} is devoted to a summary of the paper.

\section{An extension with a real singlet scalar}
%
\subsection{A model}
%
The scotogenic model proposed in \cite{ma} is an extension of the SM with an inert doublet scalar $\eta$ and right-handed neutrinos $N_k$. While they are assumed to have odd parity under the $Z_2$ symmetry, all the contents of the SM are assigned its even parity. Thus, the model is characterized by the following $Z_2$ invariant terms in Lagrangian,
\begin{align}
  -\mathcal{L}_O
  =&\sum_{\alpha,k=1}^3 \left(h_{\alpha k} \bar{\ell}_\alpha \eta N_k
  + \frac{M_{N_k}}{2} \bar{N}_k^c N_k
  + \mathrm{H.c.}\right) \nonumber \\
  &+ m_\phi^2 \phi^\dagger \phi + m_\eta^2 \eta^\dagger \eta
  + \lambda_1 (\phi^\dagger \phi)^2 + \lambda_2 (\eta^\dagger \eta)^2
  + \lambda_3 (\phi^\dagger \phi) (\eta^\dagger \eta)
  + \lambda_4 (\phi^\dagger \eta) (\eta^\dagger \phi) \nonumber \\
  &+ \frac{\lambda_5}{2} \left[(\phi^\dagger \eta)^2 + \mathrm{H.c.}\right],
  \label{model0}
\end{align}
where $\ell_\alpha$ is a doublet lepton and $\phi$ is an ordinary doublet Higgs scalar. Since $\eta$ is supposed to have no vacuum expectation value (VEV), $Z_2$ is kept as an exact symmetry of the model. As a result, the lightest neutral $Z_2$ odd field is stable to be DM. Among them, the lightest neutral component of $\eta$ is known to be a good DM candidate which does not cause any contradiction with known experimental data as long as its mass is in the TeV range \cite{ks}. On the other hand, although neutrinos cannot get mass at tree-level by the $Z_2$ symmetry, neutrino masses could be generated through a one-loop diagram and their formula is given as
\begin{align}
  \mathcal{M}^\nu_{\alpha\beta}
  \simeq \sum_{k=1}^3 h_{\alpha k}^\ast h_{\beta k}^\ast
  \left[\frac{\lambda_5 \langle\phi\rangle^2}{8 \pi^2 M_{N_k}}
  \frac{M_{N_k}^2}{M_\eta^2 - M_{N_k}^2}
  \left(1 + \frac{M_{N_k}^2}{M_\eta^2 - M_{N_k}^2}\ln\frac{M_{N_k}^2}{M_\eta^2}\right)
  \right],
  \label{nmass}
\end{align}
where $M_\eta^2 = m_\eta^2 + (\lambda_3 + \lambda_4) \langle\phi\rangle^2$. This formula suggests that small neutrino mass could be obtained for TeV scale $M_{N_k}$ and $M_\eta$ as long as $|\lambda_5|$ is small enough. Especially, it should be noted that $M_{N_k}$ can take TeV scale values to realize the required neutrino mass even if extremely small neutrino Yukawa couplings $h_{\alpha k}$ are not assumed.

In this original model, mass terms of $N_k$ are introduced by hand. Here we replace them with Yukawa couplings with a real singlet scalar $S$ which is introduced additionally. The mass term of $N_k$ is induced through these couplings if $S$ gets a VEV  $\langle S\rangle$ at a certain scale. In this extension, we also enlarge the discrete symmetry $Z_2$ to $Z_4$, under which $S$, $\eta$ and $N_k$ are supposed to have charge 2, 1 and $ -1$, respectively. All the SM contents are assumed to have no charge of it. The extended model is fixed by a $Z_4$ invariant Lagrangian and its relevant parts for the new fields are given as
\begin{align}
  -\mathcal{L}
  =&\sum_{\alpha,k=1}^3 \left(h_{\alpha k} \bar{\ell}_\alpha \eta N_k
  + \frac{y_k}{2} S \bar{N}_k^c N_k
  + \mathrm{H.c.}\right) \nonumber \\
  &+ \tilde{m}_\phi^2 \phi^\dagger \phi + \tilde{m}_\eta^2 \eta^\dagger \eta
  + \lambda_1 (\phi^\dagger \phi)^2 + \lambda_2 (\eta^\dagger \eta)^2
  + \lambda_3 (\phi^\dagger \phi) (\eta^\dagger \eta)
  + \lambda_4 (\eta^\dagger \phi) (\phi^\dagger \eta) \nonumber \\
  &+ \frac{\tilde{\lambda}_5}{2}\frac{S}{\Lambda}
  \left[(\phi^\dagger \eta)^2 + \mathrm{H.c.}\right]
  + \frac{m_S^2}{2} S^2 + \frac{\kappa_1}{4} S^4
  + \frac{\kappa_2}{2} S^2 \eta^\dagger \eta
  + \frac{\kappa_3}{2} S^2 \phi^\dagger \phi,
  \label{model}
\end{align}
where $\Lambda$ is a cutoff scale of the model. It is assumed to satisfy $\langle S\rangle \ll \Lambda$.\footnote{Since all terms invariant under the imposed symmetry up to dimension 5 are listed, effects of the cutoff scale appear only through a $\tilde\lambda_5$ term at this level.} Since $\langle S\rangle$ breaks the discrete symmetry $Z_4$ to $Z_2$, the symmetry structure is the same as the one in the original model after this breaking. Even in that case, this extension does not change the neutrino mass formula \eqref{nmass} since a term $S^2\eta^2$ is forbidden in Eq.~\eqref{model}. If we define the fluctuation $\tilde{s}$ around the vacuum after the symmetry breaking such as $S = \langle S\rangle + \tilde{s}$, parameters in Eq.~\eqref{model0} are determined by using the ones in Eq.~\eqref{model} as
\begin{align}
  M_{N_k} = y_k \langle S\rangle, \quad
  \lambda_5 = \tilde{\lambda}_5 \frac{\langle S\rangle}{\Lambda}, \quad
  m_\eta^2 = \tilde{m}_\eta^2 + \frac{\kappa_2}{2} \langle S\rangle^2, \quad
  m_\phi^2 = \tilde{m}_\phi^2 + \frac{\kappa_3}{2} \langle S\rangle^2.
  \label{cpara}
\end{align}
The mass of $\tilde{s}$ is fixed as $m_{\tilde{s}}^2 = 2 \kappa_1 \langle S\rangle^2$. Since $\langle S\rangle$ is supposed to be much larger than the weak scale here, couplings of $S$ with other scalars are assumed to be small enough so that we can safely neglect radiative effects by them. Since we consider a case where the mass of $\eta$ is smaller than $M_k$, the lightest neutral component of $\eta$ can be identified with DM.

The extra doublet scalar $\eta$ has four physical components, that is, charged ones $\eta^\pm$, and neutral ones $\eta^0_R$ and $\eta^0_I$ which are defined as $\eta^0 = (\eta^0_R + i \eta^0_I)/\sqrt{2}$. After the Higgs doublet $\phi$ gets a VEV, their mass is expressed as
\begin{align}
  M_{\eta^\pm}^2 = m_\eta^2 + \lambda_3 \langle\phi\rangle^2, \quad
  M_{\eta^0_R}^2 = m_\eta^2 + \lambda_+ \langle\phi\rangle^2, \quad
  M_{\eta^0_I}^2 = m_\eta^2 + \lambda_- \langle\phi\rangle^2,
  \label{etamass}
\end{align}
where $\lambda_\pm \equiv \lambda_3 + \lambda_4 \pm \lambda_5$ is used. The mass of these components is found to be nearly degenerate as long as $m_\eta^2 \gg \langle\phi\rangle^2$ is satisfied at least. In particular, the mass difference $\delta \equiv |M_{\eta^0_I} - M_{\eta^0_R}| \simeq |\lambda_5| \langle\phi\rangle^2/M_{\eta^0_R}$ can be very small for $|\lambda_5| \ll 1$, which is expected naturally from the smallness of neutrino mass as mentioned above. It should be also noted that $\lambda_4 < 0$ is satisfied since DM has to be electrically neutral. In the following study, we suppose $\lambda_5 < 0$ and then the lightest one is $\eta^0_R$. Although coupling constants $\lambda_i$ are free parameters of the model, they have several constraints at this stage. The stability of scalar potential of $\phi$ and $\eta$ in Eq.~\eqref{model} is known to impose the conditions such as
\begin{align}
  \lambda_1,~\lambda_2 > 0, \qquad
  \lambda_3,~\lambda_\pm > - 2 \sqrt{\lambda_1 \lambda_2}.
  \label{stab}
\end{align}
If we apply the observed Higgs mass to $m_h^2 = 4 \lambda_1 \langle\phi\rangle^2$, we have $\lambda_1 \simeq 0.13$. We also have a condition $\lambda_3,~\lambda_\pm > - 0.72 \sqrt{\lambda_2}$ by applying this to the second condition in Eq.~\eqref{stab}. We also impose the perturbativity of the model, which may be expressed as $|\lambda_i |>4\pi$.

As a new feature of the model, it should be noted that there are interaction terms relevant to $\tilde{s}$, which do not exist in the original model,
\begin{align}
  \begin{split}
		&\mathrm{(i)}\quad \frac{y_k}{2} \tilde{s} \bar{N}_k^c N_k
		+ \frac{y_k}{2} \tilde{s} \bar{N}_k N_k^c, \\
	  &\mathrm{(ii)}\quad \kappa_2 \langle S\rangle \tilde{s} \eta^\dagger \eta
	  + \frac{\kappa_2}{2} \tilde{s}^2 \eta^\dagger \eta
	  + \kappa_3 \langle S\rangle \tilde{s} \phi^\dagger \phi
	  + \frac{\kappa_3}{2} \tilde{s}^2 \phi^\dagger \phi.
	\end{split}
\end{align}
First, since $\tilde{s}$ is supposed to play a role of inflaton, the interaction in (i) could contribute to reheating after inflation if $m_{\tilde{s}} > 2 M_{N_k}$ is satisfied. Even if $m_{\tilde{s}} > 2 M_{N_k}$ is not satisfied and Yukawa coupling $h_{\alpha k}$
is extremely small, a right-handed neutrino $N_k$ could be brought in the thermal equilibrium through this interaction. It could open a possibility for low scale leptogenesis. If leptogenesis is caused by out-of-equilibrium decay of a right-handed neutrino with such $h_{\alpha k}$, no contradiction could be caused among the parameters which explain the neutrino oscillation data \cite{lowlept}. Second, the ones in (ii) could affect both Higgs physics and DM physics. If $\kappa_3$ is not small, it causes a dangerous mixing between $\tilde{s}$ and the Higgs boson. In the present study, we assume $\kappa_3=0$ to escape it.\footnote{Although $\eta$-loop generates the $S^2 \phi^\dagger \phi$ coupling radiatively, $\kappa_2$ is required to be sufficiently small as discussed later so that we can escape the dangerous mixing under this assumption.} Remaining terms could change DM phenomenology. They could change the estimation of the relic abundance of $\eta^0_R$ in the original model largely since the $\eta^0_R$ pair annihilation can be mediated by $\tilde{s}$. In the following part, we focus our study on such an interesting case defined by the mass spectrum
\begin{align}
  2 M_{\eta^0_R} \simeq m_{\tilde{s}} < 2 M_{N_k}.
  \label{masspat}
\end{align}
%

\subsection{Inflation}
%
We should note that $S$ could play a role of inflaton in addition to give the origin of the right-handed neutrino masses. It has been known that a scalar field coupled with Ricci scalar can cause an exponential expansion of the Universe \cite{nonm-inf}. Applying this idea to the SM, Higgs inflation has been proposed in \cite{higgsinf} as a realistic scenario for cosmological inflation. After this proposal, the scenario has been studied from various viewpoints \cite{h-inf1}. Recent Planck data suggest that the Higgs inflation scenario is one of favored inflation models. However, if a multicomponent field like the Higgs doublet scalar is supposed to play a role of inflaton in this framework, the model could be suffering from unitarity problems \cite{unitarity1,unitarity2}. Since unitarity could be violated at a lower scale than an inflation scale through scattering amplitudes among scalars with nonminimal couplings with a Ricci scalar, new physics required for unitarity restoration could jeopardize the flatness of the inflaton potential at the inflation scale. It can be solved in a real singlet inflaton as discussed in \cite{unitarity2}. We apply this idea to $S$ in this model.\footnote{A study of Higgs inflation in the inert doublet model can be found in \cite{inert-inf}. Although the present inflation scenario and its prediction are essentially the same as the ones in \cite{higgsinf,sinfl},  the present inflaton could play a crucial role in the DM phenomenology.} We suppose that only the singlet scalar $S$ has a non-negligible nonminimal coupling with the Ricci scalar.

The action relevant to the present inflation scenario is given in Jordan frame as
\begin{align}
  S_J
  = \int d^4x \sqrt{-g} \left[-\frac{1}{2} M_{\mathrm{pl}}^2 R
  - \frac{1}{2} \xi S^2 R + \frac{1}{2} \partial^\mu S \partial_\mu S - V(S)\right],
  \label{inflag}
\end{align}
where $M_{\mathrm{pl}}$ is the reduced Planck mass and $V(S)$ stands for a corresponding part of the potential for $S$ in Eq.~\eqref{model}. We take $S$ as an inflaton and other scalars are assumed to have much smaller values than $S$ during the inflation. In that case, $V(S)$ can be approximately expressed as $V(S) \simeq \kappa_1 S^4/4$ for a sufficiently large value of $S$. In order to derive the action in Einstein frame corresponding to Eq.~\eqref{inflag}, we use a conformal transformation \cite{higgsinf,nonm-inf}
\begin{align}
  g_{\mu \nu} = \Omega^2 g^E_{\mu \nu}, \qquad
  \Omega^2 = 1 + \frac{\xi S^2}{M_{\mathrm{pl}}^2}.
\end{align}
As a result of this transformation, we find that it can be written as
\begin{align}
  S_E
  = \int d^4x \sqrt{-g^E} \left\{-\frac{1}{2} M_{\mathrm{pl}}^2 R_E
  + \frac{1}{2 \Omega^4}\left[1 + \frac{(\xi + 6 \xi^2) S^2}{M_{\mathrm{pl}}^2}\right]
  \partial^\mu S \partial_\mu S - \frac{1}{\Omega^4} V(S)\right\}.
  \label{elag}
\end{align}
If a canonically normalized field $\chi$ is introduced as
\begin{align}
  \frac{d\chi}{dS}
  = \frac{\left[1 + (\xi + 6 \xi^2) \frac{S^2}{M_{\mathrm{pl}}^2}\right]^{1/2}}
  {1 + \frac{\xi S^2}{M_{\mathrm{pl}}^2}},
  \label{chi}
\end{align}
the potential $V(S)/\Omega^4$ in Eq.~\eqref{elag} can be expressed by using this $\chi$. It is easily seen that the new field $\chi$ coincides with $S$ at a region where $S \ll M_{\mathrm{pl}}/\sqrt{\xi}$ is satisfied. On the other hand, if $S$ takes a large value such as $S \gg M_{\mathrm{pl}}/\sqrt{\xi}$, $S$ and $\chi$ are found to be related as $S \propto \exp\big(\chi/\sqrt{6 + \frac{1}{\xi}} M_{\mathrm{pl}}\big)$. The potential at this region is almost constant
\begin{align}
  V_E
  = \frac{\kappa_1 S^4}{4 \left(1 + \frac{\xi S^2}{M_{\mathrm{pl}}^2}\right)^2}
  \simeq \frac{\kappa_1 M_{\mathrm{pl}}^4}{4 \xi^2}.
  \label{infpot}
\end{align}
This suggests that $\chi$ could play a role of the slow-rolling inflaton in this region.

The number of $e$-foldings induced by the potential $V_E$ can be estimated as
\begin{align}
  N
  = \frac{1}{M_{\mathrm{pl}}^2} \int_{\chi_{\mathrm{end}}}^\chi d\chi \frac{V_E}{V_E^\prime}
  \simeq \frac{3}{4} \frac{S^2 - S_{\mathrm{end}}^2}{M_{\mathrm{pl}}^2/\xi},
  \label{efold}
\end{align}
where $V_E^{\prime} = dV_E/d\chi$ and Eq.~\eqref{chi} is used. Slow role parameters derived from this potential can be summarized as \cite{revinf}
\begin{align}
  \varepsilon
  = \frac{M_{\mathrm{pl}}^2}{2} \left(\frac{V_E^\prime}{V_E}\right)^2
  = \frac{4 M_{\mathrm{pl}}^4}{3 \xi^2 S^4}, \qquad
  \eta
  = M_{\mathrm{pl}}^2 \left(\frac{V_E^{\prime \prime}}{V_E}\right)
  = -\frac{4 M_{\mathrm{pl}}^2}{3 \xi S^2}.
  \label{slow}
\end{align}
Since the inflation is considered to end at $\varepsilon\simeq 1$, we have $S_{\mathrm{end}}^2 \simeq \sqrt{4/3} M_{\mathrm{pl}}^2/\xi$, which suggests that $S_{\mathrm{end}}$ could be neglected in Eq.~\eqref{efold}. Thus, the slow roll parameters are found to be expressed as $\varepsilon \simeq 3/(4 N^2)$ and $\eta \simeq - 1/N$ by using the $e$-foldings number $N$ only. The spectrum of density perturbation predicted by the inflation is known to be expressed as
\begin{align}
  \mathcal{P}(k) = A_s \left(\frac{k}{k_\ast}\right)^{n_s-1}, \qquad
  A_s = \left.\frac{V_E}{24 \pi^2 M_{\mathrm{pl}}^4 \varepsilon} \right|_{k_\ast}.
  \label{power}
\end{align}
If we use $A_s=(2.101^{+0.031}_{-0.034})\times 10^{-9}$ at $k_\ast = 0.05\, \mathrm{Mpc}^{-1}$ \cite{planck18}, we find that the relation
\begin{align}
  \kappa_1 \simeq 1.49 \times 10^{-6} \xi^2 N^{-2},
  \label{kap}
\end{align}
which should be satisfied at the horizon exit time of the scale $k_\ast$. The spectral index $n_s$ and the tensor-to-scalar ratio $r$ are represented by using the slow-roll parameters as \cite{revinf}
\begin{align}
  n_s = 1 - 6 \varepsilon + 2 \eta, \qquad
  r = 16 \varepsilon.
\end{align}
Using the above results in these formulas, they are found to be $n_s\sim 0.965$ and $r\sim 3.3\times 10^{-3}$ for $N=60$. These values coincide well with the ones suggested by the Planck data \cite{planck18}. Although all these results are the same as the ones found in the Higgs inflation, the quartic coupling $\kappa_1$ is a free parameter in this model. It is completely different from the Higgs inflation case where the corresponding quartic coupling $\lambda_1$ is strictly constrained by the Higgs mass 125 GeV. This fact allows $\xi$ to take a much smaller value in comparison with the one of the usual Higgs inflation. For example, $\xi=O(10^2)$ realizes the observed value of $A_s$ for $N=60$ if $\kappa_1=O(10^{-6})$ is assumed. This $\kappa_1$ value suggests that the VEV of $S$ has to take $O(10^6)$~GeV for $m_{\tilde{s}}=O(1)$~TeV.
\begin{figure}[t]
	\begin{center}
		\includegraphics[width=14cm]{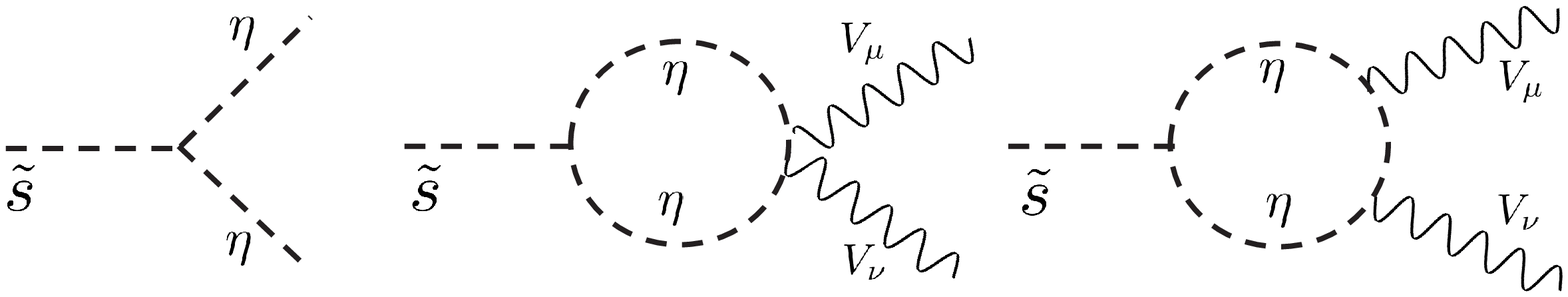}
	\end{center}
	\vspace*{-3mm}
	\footnotesize{\textbf{Fig.~1}~Feynman diagrams of dominant processes which contribute to the inflaton decay. $V_\mu$ represents gauge bosons in the SM, that is,  $W^\pm_\mu$ and $Z_\mu^0$. Higgs scalar is also allowed as the final state.}
\end{figure}

The inflaton $\tilde{s}$ starts oscillation around the vacuum $\langle S\rangle$ after the end of inflation. During this oscillation, $\tilde{s}$ is expected to decay to light fields through several modes. Since the mass pattern in Eq.~\eqref{masspat} is assumed in this study, the decay process of the inflaton is restricted to $\tilde{s} \rightarrow \eta^\dagger\eta$ at tree-level. However, $\tilde{s}$ could also decay to a pair of SM gauge bosons, Higgs bosons and neutrinos through one-loop diagrams which have $\eta$ or $N_k$ in internal lines. A part of the relevant Feynman diagrams are shown in Fig.~1. Since neutrino Yukawa  couplings and electromagnetic coupling are small compared with others, the decay to a neutrino pair and photons can be neglected among them. The decay width could be estimated as\footnote{A supplemental discussion for this derivation is given in Appendix A.}
\begin{align}
  \Gamma_{\tilde{s}}
  \simeq& \frac{(\kappa_2 \langle S\rangle)^2}{32 \pi m_{\tilde{s}}}
  \sqrt{1 - \frac{4 M_{\eta^0_R}^2}{m_{\tilde{s}}^2}} \nonumber \\
  &+ \frac{(\kappa_2 \langle S\rangle)^2}{4096 \pi^5 m_{\tilde{s}}}
  \left[\frac{(2 c_w^4 + 1) g^4}{c_w^4} \left|\mathcal{I} \left(\frac{m_\eta^2}{m_{\tilde{s}}^2}\right)\right|^2
  + \frac{1}{2} (\lambda_+ + \lambda_- + 2 \lambda_3)^2
  \left|\mathcal{J} \left(\frac{m_\eta^2}{m_{\tilde{s}}^2}\right)\right|^2\right],
  \label{width}
\end{align}
where $g$ and $\theta_W$ are a $SU(2)$ gauge coupling constant and the Weinberg angle, respectively.\footnote{In the following part, we use the abbreviation such as $c_w = \cos\theta_W$, $s_w = \sin\theta_W$ and $t_w = \tan\theta_W$.} $\mathcal{I}(r)$ and $\mathcal{J}(r)$ are defined in Eq.~\eqref{int} of Appendix A. Especially,  we should note that the one-loop contribution could become comparable with the tree-level one due to kinematic suppression for the latter if $\Delta \equiv 1 - 4 M_{\eta^0_R}^2/m_{\tilde{s}}^2 < O(10^{-4})$ is satisfied. Here we should note that the quantity in the brackets of the one-loop contribution in this $\Gamma_{\tilde{s}}$ takes a value of $O(1)$. This decay width determines reheating temperature after inflation as
\begin{align}
  T_R
  \simeq 0.53 \sqrt{M_{\mathrm{pl}}\Gamma_{\tilde{s}}}
  = O(10^{7})
  \left(\frac{\kappa_2}{\sqrt{\kappa_1}}\right)
  \left(\frac{m_{\tilde{s}}}{1\, \mathrm{TeV}}\right)^{1/2}\, \mathrm{GeV} .
\end{align}
It also depends on other model parameters $m_\eta, \lambda_\pm$ and $\lambda_3$ than $\kappa_1, \kappa_2$ and $m_{\tilde{s}}$ included in the above formula. They are constrained through DM phenomenology as discussed below. Taking account of them, expected values of $\xi$, $\langle S\rangle$ and $T_R$ are given for typical parameter sets in Table~\Rnum{1}. We note that both $T_R$ and $\langle S\rangle$ could be related to $\xi$ through Eq.~\eqref{kap} for a fixed value of $m_{\tilde{s}}$. If $T_R >\langle S\rangle$ is satisfied, the restoration of $Z_4$ could happen after the reheating and a domain wall problem could appear. However, it could be escapable by assuming a smaller value for $\kappa_2$.
\begin{figure}[t]
  \begin{center}
    \begin{tabular}{ccccc|cccc}\hline
      & $m_{\tilde{s}}${\small(GeV)} & $\kappa_1$ & $\lambda_+$ & $\lambda_3$& $\xi$ &
      $\langle S\rangle$\small{(GeV)} & $T_R$\small{(GeV)} & $Y_B$ \\ \hline
      (A) & 2000 & $10^{-6}$ & $- 0.38$ & 0.2 & 49 & $1.4\times 10^6$ &  $3.5 \times 10^5$ & $5.0 \times 10^{-11}$ \\
      (B) & 2000 & $10^{-7}$ & $- 0.38$ & 0.2 & 16 & $4.5\times 10^6$ &
      $1.1 \times 10^6$ & $9.4 \times 10^{-11}$ \\
      (C) & 2500 & $10^{-6}$ & $- 0.48$ & 0.3 & 49 & $1.8 \times 10^6$ & $3.9 \times 10^5$ & $6.3 \times 10^{-11}$ \\
      (D) & 2500 & $10^{-7}$ & $- 0.48$ & 0.3 & 16 & $5.6 \times 10^6$ & $1.2 \times 10^6$ & $1.1 \times 10^{-10}$ \\
      (E) & 3000 & $10^{-6}$ & $- 0.58$ & 0.45 & 49 & $2.1\times 10^6$ & $4.3 \times 10^5$ & $7.5 \times 10^{-11}$ \\
      (F) & 3000 & $10^{-7}$ & $- 0.58$ & 0.45 & 16 & $6.7\times 10^6$ & $1.3 \times 10^6$ & $1.3 \times 10^{-10}$ \\ \hline
    \end{tabular}
  \end{center}
  \vspace*{2mm}
  \footnotesize{\textbf{Table~\Rnum{1}}~A vacuum expectation value $\langle S\rangle$ and reheating temperature expected for assumed values of model parameters. $\Delta$ and $\kappa_2$ are fixed at $10^{-6}$ and $4 \times 10^{-6}$ in all cases, respectively. $\lambda_+$ and $\lambda_3$ are fixed by taking account of constraints from DM phenomenology discussed later.}
\end{figure}

\subsection{Leptogenesis}
%
In this model, baryon number asymmetry is expected to be generated through leptogenesis \cite{leptg}. Whether reheating temperature expected in the present inflation scenario could make thermal leptogenesis possible or not is a crucial problem for the model. As found in Table~\Rnum{1}, reheating temperature is not high enough to produce sufficient baryon number asymmetry through usual thermal leptogenesis in the original scotogenic neutrino mass model \cite{ks}. Successful leptogenesis requires much higher reheating temperature such as $T_R > 10^8$ GeV. However, in that case, both the production and the out-of equilibrium decay of right-handed neutrinos are assumed to be caused by neutrino Yukawa couplings only. Thus, the lightest right-handed neutrino is difficult to be generated in the equilibrium only by the neutrino Yukawa couplings in a consistent way with both the neutrino mass generation and the generation of sufficient lepton number asymmetry. This makes low scale leptogenesis difficult in the original model.\footnote{Low scale leptogenesis has been discussed in the scotogenic model \cite{lepta}. In these studies, the lightest right-handed neutrino is assumed to be in the thermal equilibrium through unfixed additional interaction.}

On the other hand, there is the interaction between right-handed neutrinos and the singlet scalar in this model. Since the inflation requires $\langle S\rangle = O(10^6)$ GeV for $m_{\tilde{s}} = O(10^3)$ GeV, the coupling constant $y_k$ could have a rather large value such as $O(10^{-1})$ to realize $M_{N_k} = O(10^5)$ GeV for example. This interaction could make the lightest right-handed neutrino in thermal equilibrium through the scattering mediated by the singlet scalar as long as heavier right-handed neutrinos are in the thermal equilibrium. This could occur generally even if the neutrino Yukawa coupling of the lightest right-handed neutrino is too small to make it in the thermal equilibrium. It could make successful leptogenesis possible without causing a contradiction with the neutrino oscillation data \cite{lowlept}.

In this model with a tiny $\Delta$, inflaton decays mainly to the SM gauge bosons through one-loop diagrams and then the SM contents and $\eta$ are thermalized through gauge interactions immediately. Only the right-handed neutrinos are expected to be thermalized through neutrino Yukawa couplings. Here, we remind that the neutrino oscillation data can be explained if two right-handed neutrinos have substantial Yukawa couplings $h_{\alpha k}~(k=2,3)$. An important point is that the remaining $N_1$ could be irrelevant to the neutrino mass generation. Thus, its Yukawa coupling and the mass is free from the constraints. We assume its Yukawa coupling $h_{\alpha 1}$ with doublet leptons is very small. Neutrino mass eigenvalues require $h_{\alpha k} = O(10^{-3})$ if $|\lambda_5| = O(10^{-4})$ and $M_{N_k} = 0(10^5)$ GeV are assumed. Since the decay width of $N_k$ satisfies $\Gamma_{N_k} > H(T_R)$ in such a case, $N_{2,3}$ are expected to be thermalized simultaneously at the reheating period. On the other hand, $N_1$ is expected to be thermalized through the scattering $N_k N_k \rightarrow N_1N_1$ mediated by $\tilde{s}$ since the relevant couplings have sufficient magnitude as discussed above. If $N_1$ is thermalized successfully, it decays to $\ell_\alpha \eta^\dagger$ in out-of-equilibrium through an extremely suppressed Yukawa coupling. Since the decay is delayed largely, the washout process caused by $N_k$ could be freezed out there and the generated lepton number asymmetry can be effectively converted to baryon number asymmetry through the sphaleron process. The generated lepton number asymmetry is kept escaping dilution due to the entropy production from the decay of relic $N_1$ after its substantial generation as long as the relic $N_1$ does not dominate the energy density.

We examine this scenario by solving Boltzmann equations for $Y_{N_1}$ and $Y_L (\equiv Y_\ell - Y_{\bar\ell})$, which are defined by using $f$ number density $n_f$ and entropy density $s$ as $Y_f = n_f/s$. An equilibrium value of $Y_f$ is represented by $Y_f^{\mathrm{eq}}$. As an initial condition, we assume $Y_{N_1} = Y_L = 0$ and $N_k$ is in the thermal equilibrium at $T_R$. The Boltzmann equations analyzed here are given as
\begin{align}
  \frac{dY_{N_1}}{dz}
  &= -\frac{z}{s H(M_{N_1})}
  \left(\frac{Y_{N_1}}{Y^{\mathrm{eq}}_{N_1}} - 1\right)
  \left[\gamma_D^{N_1} + \left(\frac{Y_{N_1}}{Y^{\mathrm{eq}}_{N_1}} + 1\right)
  \sum_{k=2,3} \gamma_{N_k N_k}\right], \nonumber \\
  \frac{dY_{L}}{dz}
  &= \frac{z}{s H(M_{N_1})}
  \left[\varepsilon \left(\frac{Y_{N_1}}{Y^{\mathrm{eq}}_{N_1}} - 1\right) \gamma_D^{N_1}
  - \frac{2 Y_L}{Y_\ell^{\mathrm{eq}}}
  \left(\sum_{i=1,2,3} \frac{\gamma_D^{N_i}}{4}
  + \gamma_N^{(2)} + \gamma_N^{(13)}\right)\right],
  \label{beq}
\end{align}
where $z = M_{N_1}/T$ and $H(T)$ is the Hubble parameter at temperature $T$. $\gamma_D^{N_i}$ is a reaction density for the decay $N_i \rightarrow \ell \eta^\dagger$, and $\gamma_N^{(2,13)}$ \cite{ks} and $\gamma_{N_k N_k}$ are reaction densities for lepton number violating scattering mediated by $N_k$ and scattering $N_k N_k \rightarrow N_1 N_1$ \cite{lowlept}, respectively. Although $CP$ asymmetry $\varepsilon$ is independent of flavor structure of the neutrino Yukawa couplings $h_{\alpha k}$, reaction densities could depend on it. For concreteness and simplification, we assume tri-bimaximal mixing \cite{flavor} as a rather good zeroth order approximation such as
\begin{align}
  h_{e i} = 0,~h_{\mu i} = h_{\tau i} \equiv h_i~~(i=1,~2); \quad
  h_{e 3} = h_{\mu 3} = - h_{\tau 3} \equiv h_3.
\end{align}
For numerical study of Eq.~\eqref{beq}, we use the parameters given  in Table~\Rnum{1} and other relevant ones are fixed at\footnote{A value assumed for $|\lambda_5|$ satisfies a constraint due to the DM direct search experiments which is discussed later.}
\begin{align}
  y_1 = 10^{-2}, \quad y_2 = 6 \times 10^{-2}, \quad  y_3=10^{-1}, \quad
  |\lambda_5| = 7 \times 10^{-5}, \quad  h_1=5\times 10^{-8}.
  \label{para}
\end{align}
We note that the lightest right-handed neutrino mass is of $O(10^4)$ GeV for the adopted values of $y_1$ and $\langle S\rangle$. Neutrino Yukawa couplings $h_{2,3}$ are determined to be of  $O(10^{-3})$ by using these parameters in the neutrino mass formula \eqref{nmass} and imposing neutrino oscillation data. If we assume a maximum $CP$ phase in the $CP$ asymmetry $\varepsilon$, it takes a value of $O(10^{-7})$ for the present parameter setting.

\begin{figure}[t]
	\begin{center}
		\includegraphics[width=7cm]{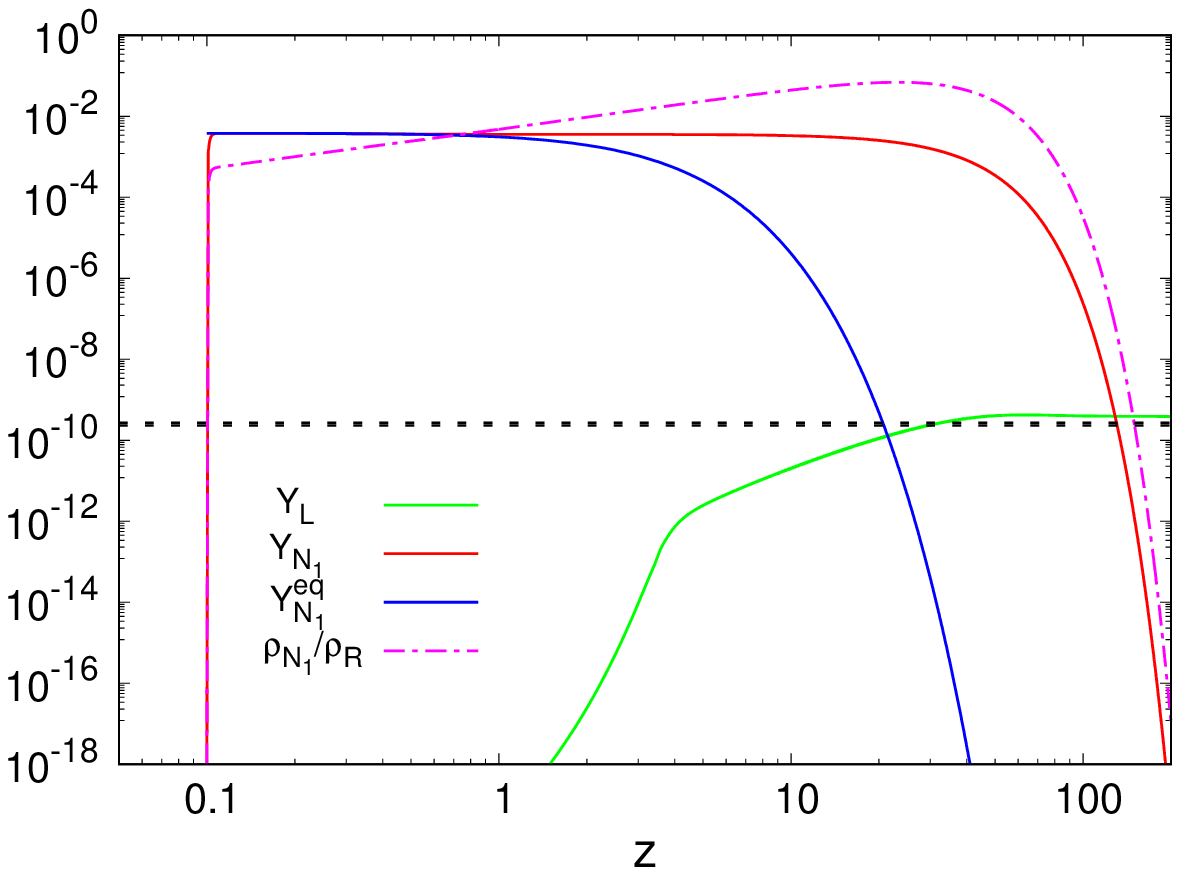}
	\end{center}
	\vspace*{-3mm}
	\footnotesize{\textbf{Fig.~2}~Evolution of $Y_{N_1}$ and $Y_L$ is shown as a function of $z$. Horizontal dotted lines show a required value of $|Y_L|$ to realize the baryon number asymmetry in the Universe $Y_B=(8.2-9.2)\times 10^{-11}$ (95\% C.L.) \cite{pdg} through  the sphaleron process. Parameters given in (F) of Table~\Rnum{1} are used in this calculation. Although the calculation is started at $z=0.1$ which are larger than $z_R$ corresponding to the reheating temperature, the result does not depend on it. A ratio $\rho_{N_1}/\rho_R$ of the $N_1$ energy density to the radiation energy density is also plotted by a dash-dotted line.}
\end{figure}

An example of solutions for the Boltzmann equations is shown in Fig.~2 to confirm the present scenario. The figure shows that the out-of-equilibrium decay starts  at $z \sim  1$ and the lepton number asymmetry is generated after it substantially. Sufficient lepton number asymmetry is found to be produced before the sphaleron decoupling at $z_{EW} \sim M_{N_1}/(10^2\, \mathrm{GeV})$. Although the $N_1$ decay is delayed, the entropy produced through the decay of $N_1$ after the substantial generation of lepton number asymmetry does not dilute it since the relic $N_1$ never dominates the energy density there. In the last column of Table~\Rnum{1}, baryon number asymmetry generated for the assumed parameters are presented. It shows that the model with suitable parameters can generate sufficient amount of baryon number asymmetry through leptogenesis although the reheating temperature is rather lower compared with the one required for successful leptogenesis in the original scotogenic model. Since the right-handed neutrino mass is generated through $M_{N_k} = y_k \langle S\rangle$, neutrino Yukawa couplings $h_{\alpha k}$ change their values under the constraints of neutrino oscillation data. The difference of $Y_B$ among the cases shown in Table~\Rnum{1} is caused by this reason. Since $M_{N_1}$ is irrelevant to the neutrino mass for the parameters in \eqref{para}, $y_1$ is free from the constraint. A larger $y_1$ can make the $CP$ asymmetry $\varepsilon$ larger without enhancing the washout effect since neutrino Yukawa couplings $h_{2,3}$ are unaffected for such a change. It suggests that $Y_B$ values shown in Table~\Rnum{1} can be made larger by assuming a larger value of $y_1$ within a region such that it makes the scattering $N_k N_k \rightarrow N_1 N_1$ decouple before $z \sim 1$.

\subsection{Inert doublet DM}
%
Here we focus our attention on the scalar $\eta$ which contains a DM candidate. The $\eta^0_R$ has several interesting features which could affect DM phenomenology as noted before. It has interaction terms
\begin{align}
  \mathcal{L}
  \supset& -\frac{\lambda_2}{4} (\eta^0_R)^4
  - \frac{\lambda_2}{2} (\eta^0_R)^2 (\eta^0_I)^2
  - \lambda_2 (\eta^0_R)^2 \eta^+ \eta^-
  - \frac{\lambda_+}{\sqrt{2}} \langle\phi\rangle h (\eta^0_R)^2
  - \frac{\lambda_+}{4} h^2 (\eta^0_R)^2 \nonumber \\
  &- \frac{\kappa_2}{2} \langle S\rangle \tilde{s} (\eta^0_R)^2
  - \frac{\kappa_2}{4} \tilde{s}^2 (\eta^0_R)^2
  + \frac{g}{2 c_w} Z_\mu
  (\eta^0_I \partial^\mu \eta^0_R - \eta^0_R \partial^\mu \eta^0_I)
  + \frac{g^2}{8 c_w^2} Z_\mu Z^\mu (\eta^0_R)^2 \nonumber \\
  &+ \frac{g^2}{4} W^+_\mu W^{- \mu} (\eta^0_R)^2
  + \frac{i g}{2} W^+_\mu (\eta^0_R \partial^\mu \eta^- - \eta^- \partial^\mu \eta^0_R)
  + \frac{i g}{2} W^-_\mu (\eta^+ \partial^\mu \eta^0_R - \eta^0_R \partial^\mu \eta^+) \nonumber \\
  &+ \frac{e g t_w}{2} Z_\mu (W^{+ \mu} \eta^- + W^{- \mu} \eta^+) \eta^0_R
  - \frac{e g}{2} A_\mu (W^{+ \mu} \eta^- + W^{- \mu} \eta^+) \eta^0_R.
\end{align}
where the physical Higgs scalar is represented by $h$. These interactions induce several processes for $\eta^0_R$. In the following part, these may be denoted as $(\eta^0_R, \eta^0_I, \eta^+, \eta^-) = (\eta_1, \eta_2, \eta_3, \eta_4)$ in some cases.

A first example is the pair annihilation of $\eta^0_R$ to a pair of $W^\pm$, $Z$s and Higgs bosons, which determines its relic abundance as DM in the Universe. Noting that total energy in the center of mass system which can be expressed as $s \simeq 4 M_{\eta_1}^2 (1 + v^2/4)$ by using relative velocity $v$ of $\eta^0_R$s, a dominant part of their pair annihilation cross section $\sigma_A v$ near the resonance $s \simeq 4 M_{\eta_1}$ is found to be given by using $\Gamma_{\tilde{s}}$ given in Eq.~\eqref{width} and $\Delta$ defined in the previous part as
\begin{align}
  \sigma_A v
  &\simeq \frac{(2 c_w^4 + 1) g^4}{128 \pi c_w^4 M_{\eta_1}^2}
  \left(1 + \mathcal{A}(s, m_{\tilde{s}}^2)\right)
  + \frac{1}{64 \pi M_{\eta_1}^2}
  \left(\lambda_+^2 + \lambda_-^2 + 2 \lambda_3^2
  + \mathcal{B}(s, m_{\tilde{s}}^2)\right),
   \nonumber \\
  \mathcal{A}(s, m_{\tilde{s}}^2)
  &= \left(\frac{\kappa_2 \langle S\rangle}{4 \pi m_{\tilde{s}}}\right)^4
  \frac{4}{(\Delta - \frac{v^2}{4})^2 + \gamma_{\tilde{s}}^2}
  \left|\mathcal{I}\left(\frac{M_{\eta_1}^2}{s}\right)\right|^2, \nonumber \\
  \mathcal{B}(s, m_{\tilde{s}}^2)
  &= \left(\frac{\kappa_2 \langle S\rangle}{4 \pi m_{\tilde{s}}}\right)^4
  \frac{(\lambda_+ + \lambda_- + 2 \lambda_3)^2}
  {(\Delta - \frac{v^2}{4})^2 + \gamma_{\tilde{s}}^2}
  \left|\mathcal{J}\left(\frac{M_{\eta_1}^2}{s}\right)\right|^2,
  \label{cross0}
\end{align}
where $\gamma_{\tilde{s}} = \Gamma_{\tilde{s}}/m_{\tilde{s}}$ is used. In this expression, we neglect contributions such as  tree-level annihilation to neutrinos, one-loop process induced by the quartic coupling $\lambda_2$, and cross terms between tree and one-loop amplitudes and so on. Nontrivial velocity dependence appears in this $\sigma_A v$ from the $s$-channel process which is mediated by an $\tilde{s}$ exchange. It could induce a crucial effect through the resonance around $v = 2 \sqrt{\Delta}$.\footnote{In various models, Breit-Wigner resonance has been extensively studied in the DM annihilation \cite{dmannih} and the DM self-interaction \cite{dmsi}.} If the velocity distribution of $\eta^0_R$ is assumed to be the Maxwell distribution $f(v)$ with velocity dispersion $\bar{v}$ which is defined through $M_{\eta_1} \bar{v}^2/2 = 3 T/2$ the velocity dependent part can be approximately averaged under a narrow resonance condition $\Delta \gg \gamma_{\tilde{s}}$ as
\begin{align}
  \int_0^\infty
  \frac{f(v)}{(\Delta - \frac{v^2}{4})^2 + \gamma_{\tilde{s}}^2}
  \simeq \frac{4 \sqrt{2} \Delta}{\sqrt{\pi}} x^{3/2} e^{- 2 \Delta x}
  \int_{- \nu_0}^{\nu_0} d\nu \frac{1}{\Delta \nu^2 + \gamma_{\tilde{s}}^2}
  = 2 \sqrt{2 \pi} x^{3/2} e^{- 2 \Delta x} \frac{\sqrt{\Delta}}{\gamma_{\tilde{s}}},
  \label{resc}
\end{align}
where we define $x \equiv M_{\eta_1}/T = 3/\bar{v}^2$ and $\nu \equiv v - 2 \sqrt{\Delta}$. Since the condition $\Delta \gg \gamma_{\tilde{s}}$ can be expressed as
\begin{align}
  \kappa_2
  \ll 10^{-3} \left(\frac{\Delta}{10^{-6}}\right)^{1/2}
  \left(\frac{\kappa_1}{10^{-6}}\right)^{1/2},
  \label{narrowg}
\end{align}
Eq.~\eqref{resc} is justified only for the case where $\kappa_1$ and $\kappa_2$ satisfy it. If it is not satisfied, small $v$
contributes substantially to the integration for large $x$ regions.
It should be corrected suitably in that case. If we use Eqs.~\eqref{cross0} and \eqref{resc}, the $\eta^0_R$ annihilation cross section averaged over the DM velocity distribution $\langle\sigma_A v\rangle$ is found to be proportional to $\left(\kappa_2/\sqrt{\kappa_1}\right)^2$ at the resonance region.

\begin{figure}[t]
  \begin{center}
    \includegraphics[width=7cm]{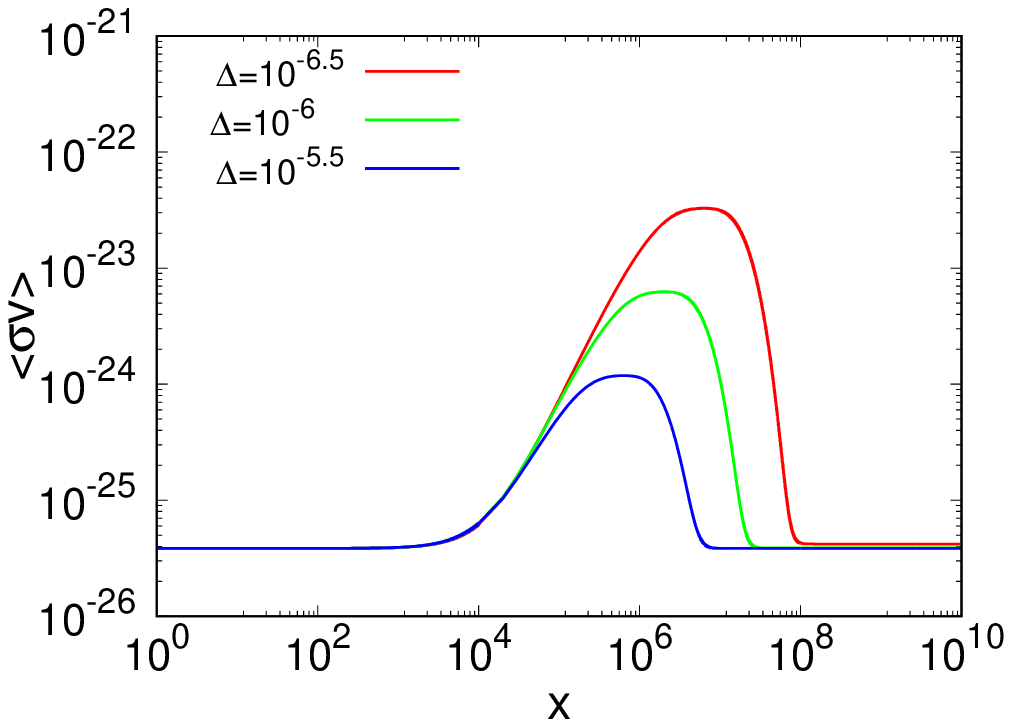}
    \hspace*{5mm}
    \includegraphics[width=7cm]{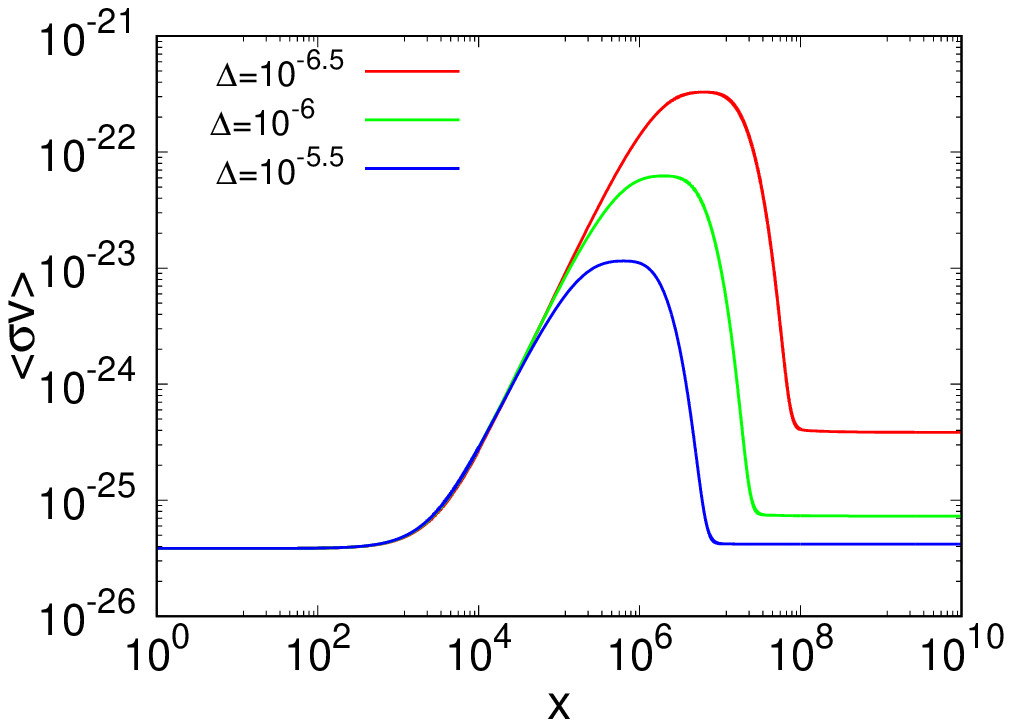}
  \end{center}
  \vspace*{-3mm}
  \footnotesize{\textbf{Fig.~3}~$\eta_R^0$ annihilation cross section $\langle\sigma_A v\rangle$ averaged over the velocity distribution with dispersion $\bar{v}$, where $x$ is related with $\bar{v}$ by $x = 3/\bar{v}^2$. $(\kappa_1, \kappa_2) $ are fixed at $(10^{-6}, 4 \times 10^{-6})$ in the left panel and $(10^{-7}, 4 \times 10^{-6})$ in the right panel. Its unit is taken as cm$^3$s$^{-1}$ in this plot. Each line corresponds to $\Delta = 10^{-6.5}$ (red),  $\Delta = 10^{-6}$ (green), and $\Delta = 10^{-5.5}$ (blue) in both panels. Other relevant model parameters are fixed at $M_{\eta_1} = 1000\, \mathrm{GeV},~\lambda_+ = -0.38,~\lambda_3 = 0.2$, and $\lambda_5=-10^{-4}$.}
\end{figure}

To find the behavior of $\langle\sigma_A v\rangle$, we plot it as a function of $x$ in Fig.~3. Here we note that the DM velocity dispersion is considered to be $\bar{v} \simeq 0.2 c$ at the freeze-out period of DM from the thermal plasma and  $\bar{v} \simeq 5 \times 10^{-5} c$ at the core of the Sun. The figure shows that the annihilation cross section of $\eta^0_R$ could have the similar value at $x$ corresponding to both velocity dispersions although the averaged cross section $\langle\sigma_A v\rangle$ has velocity dependence.\footnote{In a different context, the similar feature has been applied to the DM phenomenology in the scotogenic neutrino mass model \cite{sty}.} This might play an important role in the DM self-capture in the Sun. On the other hand, on the final relic abundance of $\eta^0_R$ in the Universe, we should note that it is not determined only by this annihilation cross section. Since the mass of the components of $\eta$ is nearly degenerate, coannihilation among all the components of $\eta$ could play a crucial role for it \cite{ks,highmass}. It suggests a possibility that the $\eta^0_R$ annihilation cross section presented above may not be directly related to its relic abundance in the model.

Elastic $\eta^0_R$-nucleon ($N$) scattering $\eta^0_R N\rightarrow \eta^0_R N$, which is relevant to the DM direct search and the DM capture in the Sun, is caused by an exchange of the Higgs boson $h$. Its cross section is given as
\begin{align}
  \sigma_N^{\mathrm{el}}
  = \frac{\lambda_+^2}{8 \pi} \frac{\bar{f}_N^2 m_N^4}{M_{\eta_1}^2 m_h^4},
  \label{elcross}
\end{align}
where $\bar{f}_N$ represents a coupling between the Higgs scalar and a nucleon. The Higgs scalar mass and the nucleon mass are represented by $m_h$ and $m_N$, respectively. Inelastic scattering $\eta^0_R N\rightarrow \eta^0_I N$ could be also brought about by a $Z$ boson exchange. Its cross section is estimated as
\begin{align}
  \sigma_N^{\mathrm{inel}} = \frac{1}{2 \pi} G_F^2 m_N^2,
  \label{inelcross}
\end{align}
where $G_F$ is the Fermi coupling constant. These could also be relevant to direct search experiments of DM. Especially, since the mass difference between $\eta^0_R$ and $\eta^0_I$ plays a crucial role in this inelastic scattering, direct DM search experiments could constrain a value of $|\lambda_5|$ as we will see it later.

Finally, we note that the $\eta^0_R$ self-scattering process such as $\eta^0_R \eta^0_R \rightarrow \eta^0_R \eta^0_R$ and $\eta^0_R \eta^0_R \rightarrow \eta^0_I \eta^0_I$ could have an influence on the DM phenomenology. In fact, the capture rate of $\eta^0_R$ in the Sun could be affected by them. The cross section of the former is calculated  as
\begin{align}
  \sigma_{RR}
  = \frac{1}{32 \pi M_{\eta_1}^2} \left|3 \lambda_2
  - \frac{\lambda_+^2}{\lambda_1}
  + \frac{(\kappa_2 \langle S\rangle)^2}{2 m_{\tilde{s}}^2}
  \left(2
  - \frac{m_{\tilde{s}}^2}{s - m_{\tilde{s}}^2 + i \Gamma_{\tilde{s}}m_{\tilde{s}}}
  \right)\right|^2,
  \label{rr}
\end{align}
where $\Gamma_{\tilde{s}}$ is the decay width of $\tilde{s}$ which is given in Eq.~\eqref{width}. It is dominated by the last term near the resonance $s \simeq m_{\tilde{s}}^2$ and behaves as
\begin{align}
  \sigma_{RR}
  \simeq \frac{1}{128 \pi M_{\eta_1}^2} \left[\frac{1}{4096 \pi^5}
  \left\{\frac{(2 c_W^4 + 1) g^4}{c_W^4}
  \left|\mathcal{I}\left(\frac{M_{\eta_1}^2}{m_{\tilde{s}}^2}\right)\right|^2
  + \frac{1}{2} (\lambda_+ + \lambda_- + 2 \lambda_3)^2
  \left|\mathcal{J}\left(\frac{M_{\eta_1}^2}{m_{\tilde{s}}^2}\right)\right|^2\right\}
  \right]^{-2}.
  \label{res}
\end{align}
Since $\left|\mathcal{I}(M_{\eta_1}^2/m_{\tilde{s}^2})\right|^2 \sim  (1 - \pi^2/4)^2$ is satisfied near the resonance, $\sigma_{RR}$ could take an enhanced value. As an example, if we suppose a case such as $\lambda_+ + \lambda_- + 2 \lambda_3 = 0$ for which $\sigma_{RR}$ is expected to take a maximum value, we find
\begin{align}
  \sigma_{RR}
  \simeq O(10^{-25}) \left(\frac{1\, \mathrm{TeV}}{M_{\eta_1}}\right)^2\, \mathrm{cm^2},
  \label{resonance}
\end{align}
which is much larger than a typical off-resonance value of $O(10^{-35})$~cm$^2$ expected for $\lambda_i = O(1)$ and $M_{\eta_1} = 1$ TeV. The most stringent constraint on the DM self-scattering cross section $\sigma$ comes from a bullet cluster \cite{bc}, which is given as $\sigma/m_{DM} \lesssim 7.0 \times 10^{-25}\, \mathrm{cm^2 GeV^{-1}}$ for DM with mass $m_{DM}$ \cite{bc,c-selfint}. It can be satisfied in the present model easily. The enhanced value of $\sigma_{RR}$ makes us expect that the self-interaction could cause a non-negligible additional contribution to the capture rate of $\eta^0_R$ in the Sun. On the other hand, the contribution from the inelastic scattering is considered to be kinematically neglected for $\eta^0_R$ at the core of the Sun. Its kinematical condition $s \geq 4 M_{\eta_2}^2$ is expressed as $\delta < M_{\eta_1} v^2/16$, where $v$ is the relative velocity. If we assume $M_{\eta_1} = O(1)$ TeV, it requires $\delta < O(10)$ keV for $\bar{v} \sim 0.7 \times 10^{-3} c$ which is expected for the scattering between $\eta^0_R$ in the Sun and $\eta^0_R$ in the Galactic halo. We find that it contradicts the result of the present direct search experiments as discussed in the next part. Thus, the inelastic scattering $\eta^0_R \eta^0_R \rightarrow \eta^0_I \eta^0_I$ is safely neglected in the estimation of the capture rate of $\eta^0_R$ in the Sun. Applying these features of $\eta^0_R$ to the experimental data for the relic DM abundance and the direct DM search, we can derive several constraints on model parameters relevant to this DM candidate.

\section{Phenomenological consequences of the singlet scalar}
%
\subsection{Experimental constraints}
%
\subsubsection{Relic abundance of $\eta^0_R$}
In the present framework, it is natural to consider that all the components of $\eta$ have the almost degenerate mass as discussed in the previous part. Thus, the $\eta^0_R$ abundance in the present Universe should be estimated taking account of the coannihilation process induced by the interactions among the components of $\eta$ in addition to Eq.~\eqref{cross0}. Its relic abundance is approximately estimated by using the formula \cite{coann}
\begin{align}
  \Omega h^2
  \simeq \frac{1.07 \times 10^9{\mathrm{GeV}}^{-1}}{J(x_F) g_\ast^{1/2} m_{\mathrm{pl}}}
\end{align}
where $m_{\mathrm{pl}}$ is the Planck mass. Freeze-out temperature $T_F(\equiv M_{\eta^0_R}/x_F)$ and $J(x_F)$ are respectively defined as
\begin{align}
  x_F
  = \ln\frac{0.0038 m_{\mathrm{pl}} g_{\mathrm{eff}} M_{\eta^0_R}
  \langle\sigma_{\mathrm{eff}} v\rangle}{(g_\ast x_F)^{1/2}}, \qquad
  J(x_F) = \int_{x_F}^\infty \frac{\langle\sigma_{\mathrm{eff}} v\rangle}{x^2} dx.
\end{align}
Effective annihilation cross section $\langle\sigma_{\mathrm{eff}} v\rangle$ and effective degrees of freedom $g_{\mathrm{eff}}$ are expressed by using the thermally averaged (co)annihilation cross section $\langle\sigma_{ij}v\rangle$ and the $\eta_i$ equilibrium number density $n_i^{\mathrm{eq}}=\left(\frac{M_{\eta_i}T}{2\pi}\right)^{3/2}e^{-\frac{M_{\eta_i}}{T}}$  as
\begin{align}
  \langle\sigma_{\mathrm{eff}} v\rangle
  = \frac{1}{g_{\mathrm{eff}}} \sum_{i,j=1}^4 \langle\sigma_{ij} v\rangle
  \frac{n_i^{\mathrm{eq}}}{n_1^{\mathrm{eq}}}
  \frac{n_j^{\mathrm{eq}}}{n_1^{\mathrm{eq}}}, \qquad
  g_{\mathrm{eff}} = \sum_{i=1}^4 \frac{n_i^{\mathrm{eq}}}{n_1^{\mathrm{eq}}}.
\end{align}

Thermally averaged (co)annihilation cross section may be expanded by the thermally averaged relative velocity $\langle v^2\rangle$ of the annihilating fields as $\langle\sigma_{ij} v\rangle = a_{ij} + b_{ij} \langle v^2\rangle$. Since $\langle v^2\rangle \ll 1$ is satisfied for cold DM, $a_{ij}$ gives dominant contribution. The effective annihilation cross section $a_{\mathrm{eff}} \equiv \sum_{i,j=1}^4 a_{ij} N_{ij}$ which is caused by both the weak gauge interactions and the quartic scalar couplings $\lambda_i$ is calculated as \cite{ks,idm}
\begin{align}
  a_{\mathrm{eff}}
  = &\frac{(1 + 2 c_w^4) g^4}{128 \pi c_w^4 M_{\eta_1}^2}
  \left(1 + \langle\mathcal{A}(s, m_{\tilde{s}}^2)\rangle\right)
  (N_{11} + N_{22} + 2 N_{34}) \nonumber \\
  &+ \frac{s_w^2 g^4}{32 \pi c_w^2 M_{\eta^0_R}^2}
  (N_{13} + N_{14} + N_{23} + N_{24}) \nonumber \\
  &+ \frac{1}{64 \pi M_{\eta_1}^2} \left[\left\{\lambda_+^2
  + \lambda_-^2 + 2\lambda_3^2
  + \langle\mathcal{B}(s, m_{\tilde{s}}^2)\rangle\right\} (N_{11} + N_{22})
  + \left\{(\lambda_+ + \lambda_-)^2\right.\right. \nonumber \\
  &\left.+ 4 \lambda_3^2
  + 2 \langle\mathcal{B}(s, m_{\tilde{s}}^2)\rangle\right\} N_{34}
  + (\lambda_+ - \lambda_-)^2 (N_{33} + N_{44} + N_{12})
  + \left\{(\lambda_+ - \lambda_3)^2\right. \nonumber \\
  &\left.\left.+ (\lambda_- - \lambda_3)^2\right\}
  (N_{13} + N_{14} + N_{23} + N_{24})\right],
  \label{cross}
\end{align}
where $\langle\mathcal{A}\rangle$ and $\langle\mathcal{B}\rangle$ are averaged values of
$\mathcal{A}$ and $\mathcal{B}$ in Eq.~\eqref{cross0} over the DM velocity distribution
and $N_{ij}$ is defined as
\begin{align}
  N_{ij}
  \equiv \frac{n_i^{\mathrm{eq}}}{n_1^{\mathrm{eq}}}
  \frac{n_j^{\mathrm{eq}}}{n_1^{\mathrm{eq}}}
  = \frac{M_{\eta_i} M_{\eta_j}}{M_{\eta_1}^2}
  \exp\left[-\frac{M_{\eta_i} + M_{\eta_j} - 2 M_{\eta_1}}{T}\right].
  \label{eqfactor}
\end{align}
%

\begin{figure}[t]
  \begin{center}
    \includegraphics[width=7cm]{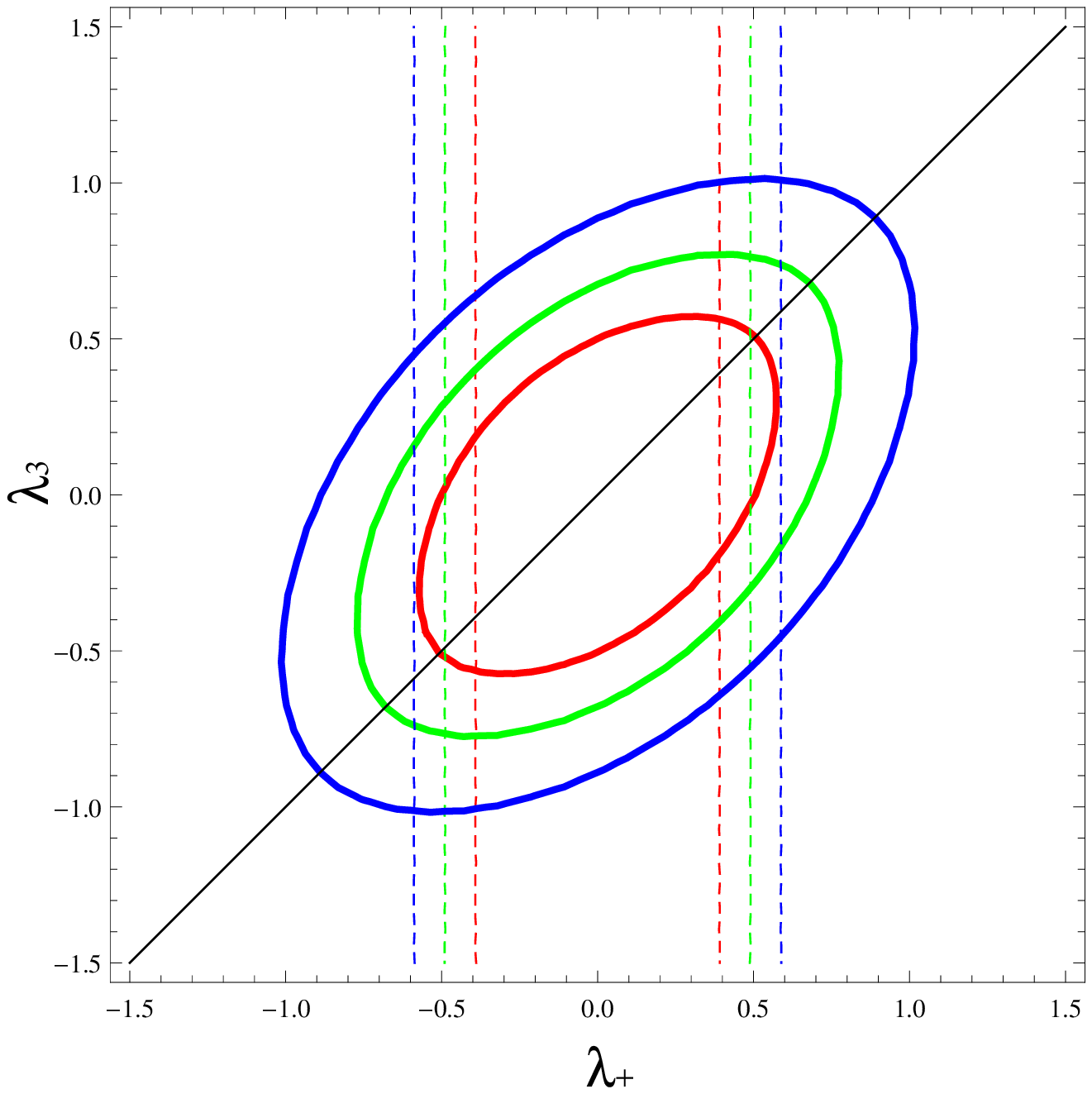}
  \end{center}
  \vspace*{-3mm}
  \footnotesize{\textbf{Fig.~4}~Contours of $\Omega h^2 = 0.12$ in the $(\lambda_+, \lambda_3)$ plane for the present model with the singlet scalar $S$. They are plotted for $M_{\eta_1} = 1000$~GeV (red solid line), 1250~GeV (green solid line) and 1500~GeV (blue solid line). $\Delta = 10^{-6}$, $\kappa_1 = 10^{-6}$, $\kappa_2 = 4 \times 10^{-6}$ are assumed. The present bounds of the direct DM search for each $\eta_R^0$ mass are also plotted by the same color dashed thin lines, which corresponds to the 90$\%$ confidence upper bound for the spin independent DM-nucleon cross section. Since it depends on $|\lambda_+|$ only, the bound appears as vertical lines in this plane. Only an upper region of the black line, which represents $\lambda_4 = \lambda_+ - \lambda_3 - \lambda_5 = 0$, is allowed, since $\lambda_4$ should be negative.}
\end{figure}
\vspace*{2mm}

We estimate the relic abundance due to freeze-out of the thermal $\eta^0_R$ by using these formulas. Since the DM velocity dispersion at this freeze-out temperature is considered to be $\bar{v} \sim 0.2 c$ which corresponds to $x_F \sim 25$, the $s$-channel process is off resonance for a small $\Delta$ such as $10^{-6}$ and it does not cause a substantial effect. In that case, main relevant free parameters contained in the cross section are the DM mass $M_{\eta_1}$, the coupling constants $\lambda_+$ and $\lambda_3$.\footnote{Since we can suppose $|\lambda_5|$ is much smaller than $|\lambda_3|$ and $|\lambda_4|$ without loss of generality based on the discussion on the small neutrino mass generation, we use these two couplings as independent parameters.} We plot contours of $\Omega h^2 = 0.12$ in the $(\lambda_+, \lambda_3)$ plane for typical values of $M_{\eta_1}$ in Fig.~4. Since $M_{\eta_{3,4}} > M_{\eta_{1,2}}$ should be satisfied, the allowed region is constrained to $\lambda_4 < 0$ which corresponds to a region above a black solid line. The figure shows that the required relic abundance can be obtained easily by choosing values of $(\lambda_+, \lambda_3)$. Since we take a small value for $\kappa_2$ so as to satisfy Eq.~(\ref{narrowg}), the singlet scalar effect is negligible in the relic abundance estimation.

\subsubsection{Direct search constraint}
Nucleus-DM elastic scattering is brought about through a $t$-channel Higgs exchange in this model. It can be a target of direct search experiments of DM and they impose a constraint on the DM-nucleon scattering cross section $\sigma^{\mathrm{el}}_n$ in Eq.~\eqref{elcross}. Thus, it constrains a value of $\lambda_+$ for a fixed value of the mass of $\eta^0_R$. The most stringent bound is presented by XENON1T \cite{crbound}. It gives a constraint on spin independent DM-nucleon cross section such as $\sigma_{\mathrm{SI}} \lesssim 8.5 \times 10^{-46}\, \mathrm{cm^2}$ for $m_{DM} =$1~TeV. If we use $m_h = 125$~GeV and $\bar{f}_N = 1/3$ in Eq.~\eqref{elcross}, we can find a bound on $\lambda_+$ such as $|\lambda_+| \lesssim 0.4$ for $M_{\eta_1} =$1~TeV. Since $\lambda_+$ is also relevant to the relic abundance of $\eta^0_R$ as seen in the previous part, we have to combine them to find an allowed region for the parameters of the model.

In Fig.~4, we show this direct DM search bound for the assumed $\eta^0_R$ mass in the $(\lambda_+, \lambda_3)$ plane. The bounds are presented by dashed thin lines with different colors for each $\eta^0_R$ mass. The same color is used as the one for the relic abundance. Since $\sigma_N^{\mathrm{el}}$ is independent on $\lambda_3$ and depends only on the absolute value of $\lambda_+$, the bounds are represented as a set of symmetric vertical lines in the $(\lambda_+, \lambda_3)$ plane for a fixed $\eta^0_R$ mass. Since a region sandwiched by these lines is remained as an allowed region, only the points on the contours $\Omega h^2 = 0.12$ contained there can be accepted. From this figure, we find that $\lambda_3 < 0$ can be excluded by adding the direct DM search bound to the relic abundance condition.

We remind here that inelastic scattering $\eta^0_R N \rightarrow \eta^0_I N$ could play an important role if the mass difference $\delta = |\lambda_5| \langle\phi\rangle^2/M_{\eta_1}$ is small enough. In fact, it could occur at the similar order magnitude to the elastic scattering such as $\sigma_N^{\mathrm{inel}}/\sigma_N^{\mathrm{el}} \simeq 3 (0.1/\lambda_+)^2$, which is found from Eqs.~\eqref{elcross} and \eqref{inelcross}. Since the direct searches find no events by now, we can consider two possibilities for it. That is, they are kinematically forbidden or  they are kinematically allowed but its signature cannot be found at the present detector sensitivity. If the process is kinematically allowed, the relative velocity $v$ between DM and a target nucleus should be larger than a certain minimum value $v_{\mathrm{min}}$. Here, $v_{\mathrm{min}}$ can be estimated as \cite{inelas,l5}
\begin{align}
  v_{\mathrm{min}}
  = \frac{1}{\sqrt{2 m_{\mathcal{N}} E_R}}
  \left(\frac{m_{\mathcal{N}} E_R}{m_r} + \delta\right),
  \label{vmin}
\end{align}
where $m_\mathcal{N}$ and $E_R$ are mass and recoil energy of a target nucleus $\mathcal{N}$ respectively, and $m_r$ is the reduced mass of $\mathcal{N}$ and DM. Since $v$ should satisfy $v < v_{\mathrm{esc}} + v_0$ where $v_{\mathrm{esc}}$ is the escape velocity from our Galaxy at the Earth ($v_{\mathrm{esc}} \simeq 544$~km/s) and $v_0$ is the circular velocity around the center of Galaxy ($v_0 \simeq 220$~km/s), the process is considered to be allowed kinematically for $v_{\mathrm{min}} < v < v_{\mathrm{esc}} + v_0$. If we insert a relation $v_{\mathrm{min}} = v_{\mathrm{esc}} + v_0$ in Eq.~\eqref{vmin} and take account of $\delta = |\lambda_5|\langle\phi\rangle^2/M_{\eta_1}$, we find a critical value of $\lambda_5$ as
\begin{align}
  |\lambda_5^c|
  \simeq 8 \times 10^{-6} \left(\frac{M_{\eta_1}}{10^3\, \mathrm{GeV}}\right)
  \left(\frac{M_\mathcal{N}}{10^2\, \mathrm{GeV}}\right)^{1/2}
  \left(\frac{E_R}{40\, \mathrm{keV}}\right)^{1/2}.
  \label{c-inel}
\end{align}
If $|\lambda_5| > |\lambda_5^c|$ is satisfied, the inelastic scattering is kinematically forbidden. On the other hand,  for the case $|\lambda_5| < |\lambda_5^c|$ we can consider a possibility that it is allowed kinematically but its signature is not found since the reaction rate is below the present detector sensitivity. However, present direct DM search experiments seem to have excluded such a possibility already \cite{ks-l}. Anyway, although inelastic scattering could contribute to the DM capture in the Sun in general \cite{inel-capt}, it needs not to be taken into account in the estimation of the $\eta^0_R$ capture in the present model. Here it may be useful to note that the bound on $\lambda_5$ and the value of $\langle S\rangle$ require the cutoff scale $\Lambda$ to be $\Lambda \lesssim 10^{11}$ GeV for $\tilde\lambda_5=O(1)$ through Eq.~\eqref{cpara}. It coincides with a scale of Peccei-Quinn symmetry breaking for the strong $CP$ problem \cite{strongcp}.

\subsection{High energy neutrinos caused by annihilation of $\eta^0_R$
captured in the Sun}
%
DM in the Galaxy can be captured inside the Sun through scattering with nuclei contained in the Sun if it loses energy and its velocity becomes smaller than the escape velocity at that point. If the captured DM annihilates to produce neutrinos, they could be a good target of indirect DM search \cite{cap-nu}. It may give an interesting signature of the present model, which could have enhanced self-scattering due to the possible resonance caused by the singlet scalar. We discuss this subject here.

Time evolution of the number $N$ of DM captured inside the Sun is described by the equation \cite{selfint}
\begin{align}
  \frac{dN}{dt} = C_c + C_s N - C_a N^2,
  \label{numb}
\end{align}
where $C_c$ and $C_a$ stand for capture rate of DM through the scattering with nuclei in the Sun and annihilation rate between DMs captured already in the Sun, respectively. The second term is caused by the self-scattering between DM in the halo of the Galaxy and DM captured in the Sun. Since the age of the Sun is larger than the timescale $\tau = 1/\sqrt{C_c C_a + C_s^2/4}$ for which DM annihilation and DM capture are in the equilibrium,\footnote{We can check that it is satisfied in this model for typical values of $C_c$, $C_a$ and $C_s$ presented below.} Eq.~\eqref{numb} is considered to reach a steady state. In that case, $N$ can be expressed as
\begin{align}
  N = \frac{C_s}{2 C_a} + \sqrt{\frac{C_s^2}{4 C_a^2} + \frac{C_c}{C_a}}.
  \label{dm-numb}
\end{align}
Since the annihilation rate of DM in the Sun is given by $\Gamma_A = C_a N^2/2$, it can be expressed as
\begin{align}
  \Gamma_A
  = \frac{1}{2} \left[C_c + \frac{C_s^2}{2 C_a}
  \left(1 + \sqrt{1 + \frac{4 C_a C_c}{C_s^2}}\right)\right].
  \label{gamma}
\end{align}
In a case of $C_s^2 \ll C_a C_c$ which corresponds to a case with negligible self-interaction, we have $\Gamma_A = C_c/2$. It is determined only by the capture rate $C_c$. If $C_s^2 \gg C_a C_c$ is satisfied, we have $\Gamma_A = C_s/2 C_a$ which is irrelevant to the capture rate $C_c$ on the contrary. These two limiting cases suggest that enhanced self-capture expected in the present model might affect high energy neutrino flux caused by the DM annihilation in the Sun to give a characteristic signature of the model.

The flux of neutrino $\nu_\alpha$ caused by the $\eta^0_R$ annihilation at the core of the Sun can be expressed as \cite{dm}
\begin{align}
  \frac{d\Phi_{\nu_\alpha}}{dE_\nu}
  = \frac{1}{4 \pi R^2} \Gamma_A \sum_f B_f
  \frac{dN_f^{\nu_\alpha}(E_\nu, E_{\mathrm{in}})}{dE_\nu}
  \label{flux}
\end{align}
where $R$ is the distance between the Sun and the Earth. $\Gamma_A$ is the annihilation rate of $\eta^0_R$ in the Sun and it is given by Eq.~\eqref{gamma}. $B_f$ is the branching ratio of the $\eta^0_R$ annihilation to a channel $f$ which is contained in Eq.~\eqref{cross0}. $dN_f^{\nu_\alpha}(E_\nu, E_{\mathrm{in}})/dE_\nu$ stands for the $\nu_\alpha$ spectrum at the surface of the Sun, when it is produced with energy $E_\nu$ through the channel $f$ with injection energy $E_{in}$. High energy $\nu_\mu$ and $\bar{\nu}_\mu$ from the Sun are searched by observing up-going muons at IceCube. Since no signature is observed still now, IceCube gives upper bound on the annihilation rate for relevant DM decay modes such as $W^+W^-$, $\tau \bar{\tau}$ and $b \bar{b}$ which cause high energy neutrinos finally. It could give some constraints on the model.

Now we proceed to estimate $C_a$, $C_c$ and $C_s$ in order to estimate $\Gamma_A$ in the model. First of all, we estimate $C_a$ which can be expressed as $C_a = \langle\sigma_A  v\rangle \int_0^{R_\odot} n_{\eta_1}^2(r) 4 \pi r^2 dr$ and $N = \int_0^{R_\odot} n_{\eta_1}(r) 4\pi r^2 dr$. Here
$n_{\eta_1}(r)$ is the $\eta^0_R$ number density in the Sun and $\langle\sigma_A v\rangle$ is the averaged total annihilation cross section given in Eq.~\eqref{cross0}. If we use an effective volume $V_j$ which is defined as $V_j = \int_0^{R_\odot}
n_{\eta_1}^j(r) 4 \pi r^2 dr$, it is expressed as $C_a = \langle\sigma_A v\rangle V_2/V_1^2$ \cite{effvol}. The number density of $\eta^0_R$ near the core of the Sun could be represented as $n_{\eta_1}(r) = \exp(-M_{\eta_1} \phi(r)/T_0)$ where $\phi(r)$ is the gravitational potential of the Sun and $T_0 = 1.57 \times 10^7$~K is the core temperature of the Sun. If we suppose a constant mass density near the core as $\rho_0 = 156\, \mathrm{g/cm^3}$, $\phi(r)$ can be given as $\phi(r) = 2 \pi G \rho_0 r^2/3$. In that case, we find that $V_j$ is approximately estimated as
\begin{align}
  V_j
  = \int_0^{R_\odot} e^{-\frac{j M_{\eta_1} \phi(r)}{T_0}} 4 \pi r^2 dr
  = \left(\frac{3 m_{\mathrm{pl}}^2 T_0}{2 j M_{\eta_1} \rho_0}\right)^{3/2}
  \simeq 7.30 \times 10^{25} \left(\frac{1\, \mathrm{TeV}}{j M_{\eta_1}}\right)^{3/2}\,
  \mathrm{cm}^3.
\end{align}
By using Eq.~\eqref{cross0} for $\langle\sigma_A v\rangle$, $C_a$ can be estimated as $C_a \simeq 1.8 \times 10^{-52}\, \mathrm{s^{-1}}$ if we take relevant parameters, as an example, to be $(\lambda_+, \lambda_3) = (-0.38, 0.2)$ for $M_{\eta_1} = 1$~TeV, which is contained in the allowed region shown in Fig.~4. Since $(\lambda_+, \lambda_3)$ has to be contained in a region limited by both the DM relic abundance and the DM direct search as shown in Fig.~4, expected values of $C_a$ cannot change largely from the value quoted above.
\begin{figure}[t]
  \begin{center}
    \includegraphics[width=7cm]{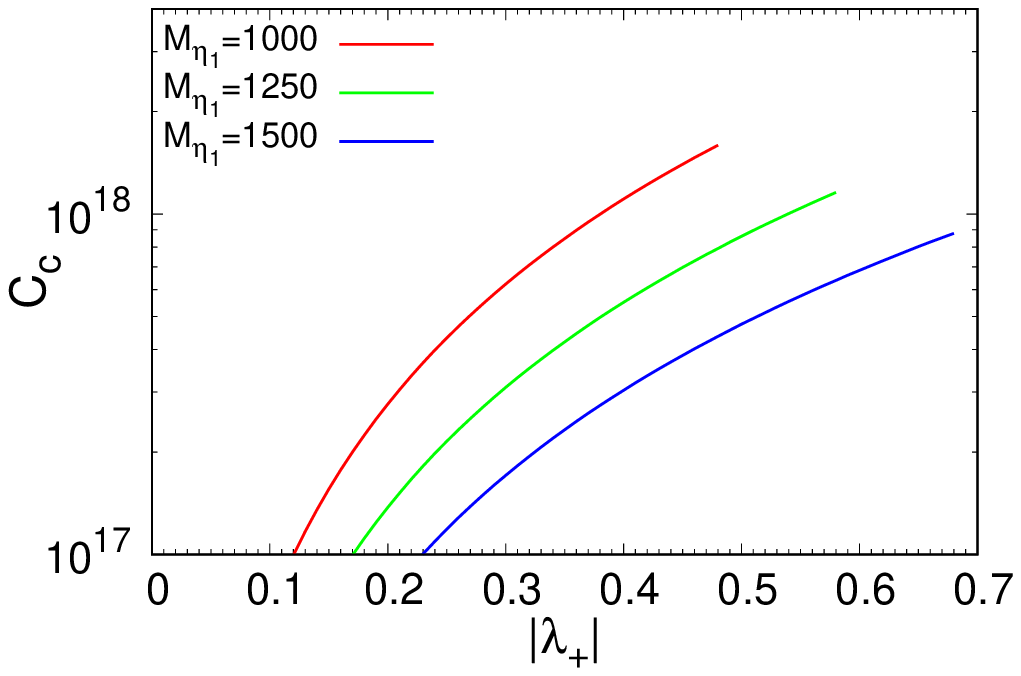}
  \end{center}
  \vspace*{-3mm}
  \footnotesize{\textbf{Fig.~5}~Capture rate $C_c\,[\mathrm{s^{-1}}]$ of $\eta^0_R$ due to $\eta^0_R$-nucleus($\mathcal{N}_i$) scattering in the Sun. It is plotted as a function of $|\lambda_+|$ for typical values of $M_{\eta_1}$. Since the upper bound of $|\lambda_+|$ is constrained by the DM relic abundance as shown Fig.~4, it is taken into account in this plot.}
\end{figure}

Next, we estimate $C_c$ and $C_s$ in the present model. For the estimation of $C_c$, we follow the argument given by Gould \cite{capt} and apply it to $\eta^0_R$ in the present model. Its expression is given by using variables defined in Appendix B as\footnote{A brief review of this derivation is given in Appendix B.}
\begin{align}
  C_c
  = &\sum_i \frac{\sigma(\eta_1 \mathcal{N}_i) \rho_{\eta_1} \bar{v} M_\odot f_i}
  {4 \sqrt 6 \zeta a_i M_{\mathcal{N}_i}^2}
  \left\{\frac{2 e^{\frac{- a_i \zeta^2}{1 + a_i}}}{\sqrt{1 + a_i}}
  \mathrm{erf} \left(\frac{\zeta}{\sqrt{1 + a_i}}\right)
  - \frac{e^{\frac{- a_i \zeta^2}{1 + a_i}}}{(A_c^2 - A_s^2)(1 + a_i)^{3/2}}\right.
   \nonumber \\
  &\times \left[\left(\hat{A}_{i+} \hat{A}_{i-} - \frac{1}{2}
  - \frac{1 + a_i}{a_i - b_i}\right) \left(\mathrm{erf}(\hat{A}_{i+})
  - \mathrm{erf}(\hat{A}_{i-})\right)\right. \nonumber \\
  &\left.+ \frac{1}{\sqrt{\pi}} \left(\hat{A}_{i-} e^{-\hat{A}_{i+}^2}
  - \hat{A}_{i+} e^{-\hat{A}_{i-}^2}\right)\right]_{A_i=A_i^s}^{A_i=A_i^c}
  + \frac{e^{\frac{-b_i \zeta^2}{1 + b_i}}}
  {(a_i - b_i)(A_c^2 - A_s^2)\sqrt{1 + b_i}} \nonumber \\
  &\left.\times \left[e^{-(a_i - b_i) A_i^2}
  \left(2\, \mathrm{erf}\left(\frac{\zeta}{\sqrt{1 + b_i}}\right)
  - \mathrm{erf}(\check{A}_{i+})
  + \mathrm{erf}(\check{A}_{i-})\right)\right]_{A_i=A_i^s}^{A_i=A_i^c}\right\}.
  \label{cc}
\end{align}
The $\eta^0_R$-nucleus($\mathcal{N}_i$) cross section $\sigma(\eta_1\mathcal{N}_i)$ in this formula is given by using $\sigma_N^{\mathrm{el}}$ in Eq.~\eqref{elcross} as
\begin{align}
  \sigma(\eta_1 \mathcal{N}_i)
  = \sigma_N^{\mathrm{el}} A_i^2
  \frac{M_{\eta_1}^2 M_{\mathcal{N}_i}^2}{(M_{\eta_1} + M_{\mathcal{N}_i})^2}
  \frac{(M_{\eta_1} + m_p)^2}{M_{\eta_1}^2 m_p^2},
  \label{cN}
\end{align}
where $m_p$ is the proton mass and $A_i$ is the atomic number of nucleus $\mathcal{N}_i$. Using this formula,  we plot $C_c$ as a function of $\lambda_+$ for several reference values of  $M_{\eta_1}$ in Fig.~5. Since $M_{\eta_1}$ is assumed to be in a TeV range which is much larger than the mass of target nucleus, the capture rate $C_c$ is kinematically suppressed to be $O(10^{18})\, \mathrm{s}^{-1}$. It suggests that $C_s$ could have substantial effects in the $\eta^0_R$ capture in the Sun only if $C_s$ takes the same order value as $\sqrt{C_a C_c} = O(10^{-17})\, \mathrm{s}^{-1}$.
\begin{figure}[t]
  \begin{center}
    \includegraphics[width=7cm]{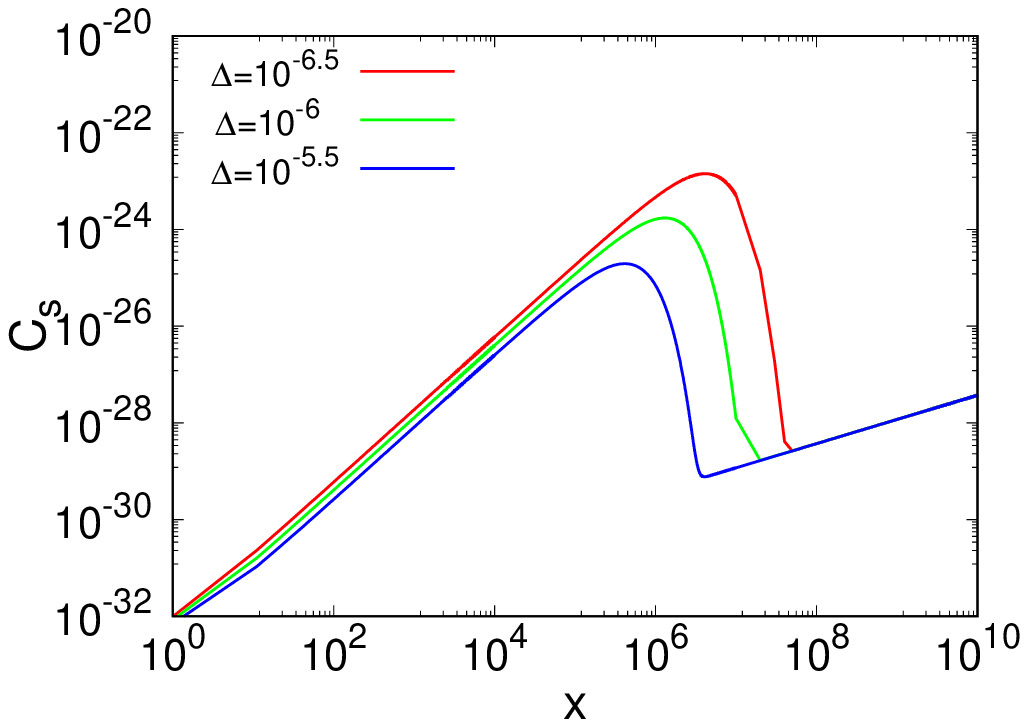}
    \hspace*{5mm}
    \includegraphics[width=7cm]{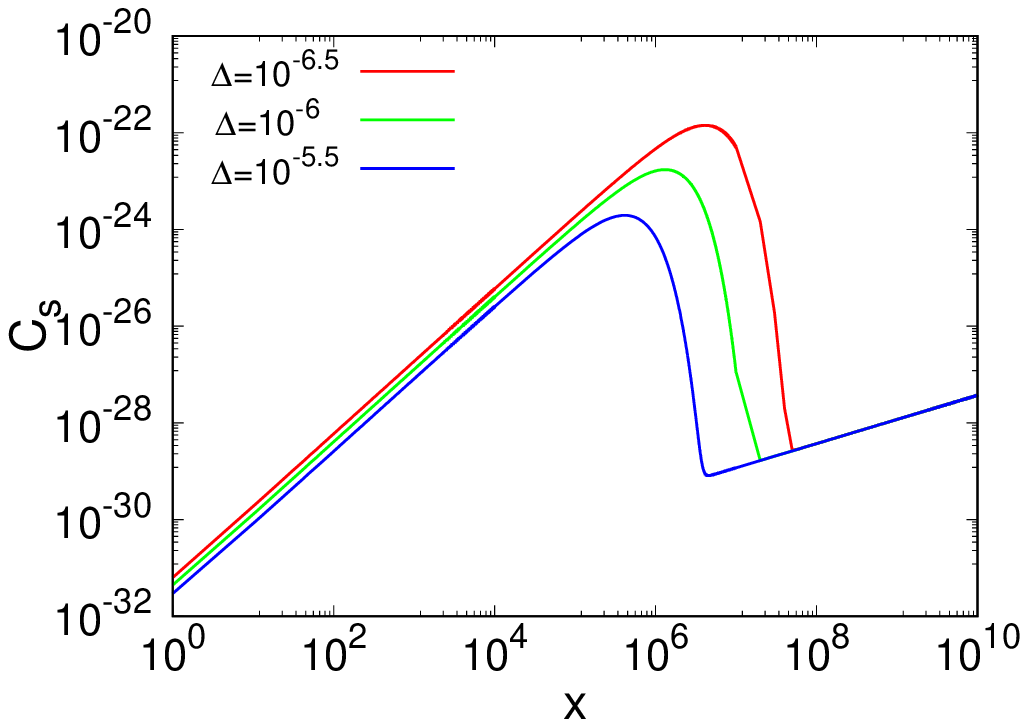}
  \end{center}
  \vspace*{-3mm}
  \footnotesize{\textbf{Fig.~6}~Capture rate $C_s\,[\mathrm{s^{-1}}]$ of $\eta^0_R$ due to the self-scattering in the Sun. It is plotted as a function of $x$. $(\kappa_1, \kappa_2)$ are fixed as $(10^{-6}, 4 \times 10^{-6})$ in the left panel and $(10^{-7}, 4 \times 10^{-6})$ in the right panel. Each line corresponds to $\Delta = 10^{-6.5}$ (red),  $\Delta = 10^{-6}$ (green), and $\Delta = 10^{-5.5}$ (blue) in both panels. Other relevant parameters are also fixed at the same values used in Fig.~3.}
\end{figure}

Self-capture rate $C_s$ can be also calculated in the same way as $C_c$ as reviewed in Appendix B. Its analytic expression is approximately obtained as
\begin{align}
  C_s
  \sim &\frac{1}{32 \pi M_{\eta_1}^2 \zeta} \frac{\rho_{\eta_1}}{M_{\eta_1}}
  \frac{v_s^2}{\bar{v}^2} \left[\sqrt{\frac{3}{2}} \bar{v}
  \left(3 \lambda_2 - \frac{\lambda_+^2}{\lambda_1}
  + \frac{2 (\kappa_2 \langle S\rangle)^2}{m_{\tilde{s}}^2}\right)^2
  \mathrm{erf}(\zeta)\right. \nonumber \\
  &+ \left.\frac{3 \sqrt{\pi}}{2} \left(e^{-\left(\frac{\sqrt{6 \Delta}}{\bar{v}}
  - \zeta\right)^2}
  - e^{-\left(\frac{\sqrt{6 \Delta}}{\bar{v}} + \zeta\right)^2}\right)
  \left(\frac{\kappa_2 \langle S\rangle}{m_{\tilde{s}}}\right)^4
  \frac{1}{\sqrt{\Delta} \gamma_{\tilde{s}}}\right]
  \left\langle\frac{v(r)^2}{v_s^2}\right\rangle,
  \label{cs}
\end{align}
where $\langle v(r)^2/v_s^2\rangle$ is a value of  the squared escape velocity averaged over the distribution of $\eta^0_R$ in the Sun. It is defined by using the number density $n(r)$ of $\eta^0_R$ in the Sun as
\begin{align}
  \left\langle\frac{v(r)^2}{v_s^2}\right\rangle
  = \frac{1}{N} \int_0^{R_\odot} 4 \pi r^2 dr n(r) \frac{v(r)^2}{v_s^2}, \qquad
  N = \int_0^{R_\odot} 4 \pi r^2 dr n(r),
  \label{escape}
\end{align}
where $v_s$ is escape velocity at the surface of the Sun. Since $\eta^0_R$ is heavy enough so as to be accumulated in the core of the Sun, the average value of $v(r)^2/v_s^2$ in Eq.~\eqref{escape} is evaluated as 5.1 \cite{capt}. In Fig.~6, we plot $C_s$ as a function of $x$ for a fixed value of $\Delta$ by using Eq.~\eqref{cs}. Since the velocity dispersion of $\eta^0_R$ in the halo is considered to be $\bar{v} \sim 10^{-3} c$, $C_s$ could be large as long as $\bar{v}^2 \simeq 2 \Delta$ is satisfied. Figures 3 and 6 suggest that $C_s$ is enhanced largely due to the resonance in the $s$-channel process caused by the $\tilde{s}$ exchange for $\Delta \simeq 10^{-6}$ while $C_a$ can keep a value required by the relic abundance there for suitable parameters because of $\bar{v} \sim 10^{-5}$ for $\eta_R^0$ in the Sun. Although the self-capture rate $C_s$ is enhanced largely, it is difficult to reach a value of $O(10^{-17})$. The enhancement is not sufficient to make the predicted value of $\Gamma_A$ deviate from $\Gamma_A = C_c/2$ substantially.

\begin{figure}[t]
  \begin{center}
    \begin{tabular}{cccc} \hline
      & $\Gamma_{WW}(\mathrm{s}^{-1})$ & (a)~$\langle\sigma_A v\rangle_{\gamma\gamma}(\mathrm{cm}^{3}\mathrm{s}^{-1})$ &
      (b)~$\langle\sigma_A v\rangle_{\gamma\gamma}(\mathrm{cm}^{3}\mathrm{s}^{-1})$ \\ \hline
      (A) & $4.2 \times 10^{17}$ & $2.3 \times 10^{-29}$ & $2.2 \times 10^{-29}$ \\
      (B) & $3.1 \times 10^{17}$ & $2.3 \times 10^{-27}$ & $2.2 \times 10^{-27}$ \\
      (C) & $2.3 \times 10^{17}$ & $1.4 \times 10^{-29}$ & $1.4 \times 10^{-29}$ \\
      (D) & $2.3 \times 10^{17}$ & $1.4 \times 10^{-27}$ & $1.4 \times 10^{-27}$ \\
      (E) & $1.8 \times 10^{17}$ & $1.0 \times 10^{-29}$ & $9.6 \times 10^{-30}$ \\
      (F) & $1.8 \times 10^{17}$ & $1.0 \times 10^{-27}$ & $9.6 \times 10^{-28}$ \\ \hline
    \end{tabular}
  \end{center}
  \vspace*{2mm}
  \footnotesize{\textbf{Table \Rnum{2}}~Predictions of the model for the the annihilation rate to $WW$, the averaged velocity weighted annihilation cross section to monochromatic gamma rays. We use typical values of the model parameter presented in Table~\Rnum{1}. The velocity dispersion is assumed to be $\bar{v} = 3 \times 10^{-4} c$ in (a) and $\bar{v} = 3 \times 10^{-5}c$ in (b).}
\end{figure}

By using $C_a$, $C_c$ and $C_s$ obtained in the above study, we can calculate the $\eta^0_R$ annihilation rate $\Gamma_A$ in Eq.~\eqref{gamma}. We consider to apply a constraint on the DM decay to $W^+ W^-$ obtained by the IceCube neutrino telescope to this model. The annihilation cross section of $\eta^0_R$s to $W^+W^-$ is given as
\begin{align}
  \langle\sigma_A v\rangle_{WW}
  = \frac{g^4}{64 \pi M_{\eta_1}^2}
  \left(1 + \langle\mathcal{A}(s, m_{\tilde{s}}^2)\rangle\right)
  + \frac{1}{128 \pi M_{\eta_1}^2} \left(4 \lambda_3^2
  + \langle\mathcal{B}(s, m_{\tilde{s}}^2)\rangle\right)
\end{align}
where $\langle\mathcal{A}\rangle$ and $\langle\mathcal{B}\rangle$ are averaged values over the DM velocity distribution in the Sun. We assume $\bar{v} \simeq 5 \times 10^{-5} c$ in this analysis. Since the relevant model parameters are $\lambda_\pm$, $\lambda_3$, $M_{\eta_1}$ and $\langle S\rangle$, we take account of the constraint on them which is found in Fig.~3 to estimate annihilation rate $\Gamma_{WW}$. The results are shown in Table~\Rnum{2}. If we compare these with the limit $\Gamma_{WW} \leq 9.34 \times 10^{19}\, \mathrm{s}^{-1}$ given by the IceCube in \cite{icecube}, the predicted values for $\Gamma_{WW}$ suggest that it is difficult to examine the model by using neutrinos generated through the $\eta_R^0$ annihilation in the Sun unless the sensitivity of experiments could be improved more than 2 orders at least.

\subsection{High energy gamma rays produced through annihilation of $\eta^0_R$}
%
It is well known that the $\eta^0_R$ annihilates to a photon pair through one-loop diagrams in the original model. However, since it is suppressed heavily, it is considered to be difficult to probe them through indirect searches. In the present model, however, it can also occur through the $s$-channel exchange of $\tilde{s}$. In that case, the cross section could be largely enhanced if the resonance condition is satisfied for the velocity dispersion of DM at certain places where the annihilation occurs. If such a situation is prepared somewhere in the Universe, monochromatic gamma rays generated through the $\eta^0_R$ annihilation there might give us an observable signature of the model in high energy gamma-ray searches such as H.E.S.S.

Gamma-ray flux $\Phi_\gamma$ caused by the $\eta^0_R$ annihilation is expressed as
\begin{align}
  \frac{d\Phi_\gamma}{dE_\gamma}
  = \frac{1}{4 \pi} \frac{\langle\sigma_A v\rangle}{2 M_{\eta_1}^2}
  \sum_f \frac{dN_\gamma^f}{dE_\gamma} B_f
  \int_{\Delta \Omega} d\Omega^\prime \int_{los}
  \rho_{\eta_1}^2(r(\ell, \phi^\prime)) d\ell(r, \phi^\prime),
  \label{g-spectrum}
\end{align}
where $B_f$ is the branching ratio to a final state $f$ which generates gamma rays and $dN_\gamma^f/dE_\gamma$ is the gamma-ray spectrum generated there. A part given by integrals represents an astrophysical factor called $J$-factor, which represents DM distribution within a solid angle $\Delta \Omega$ along a line of sight. Gamma rays produced through the $\eta^0_R$ annihilation have a line shape component. Its cross section averaged over the velocity distribution is expressed near the resonance as
\begin{align}
  \langle\sigma_A v\rangle_{\gamma \gamma}
  \simeq \frac{e^4}{32 \pi M_{\eta_1}^2} \left(\frac{16 \lambda_2^2}{(4 \pi)^4}
  + \langle\mathcal{A}(s, m_{\tilde{s}}^2)\rangle\right).
  \label{g-cross}
\end{align}
The second term comes from diagrams with the $\tilde{s}$ exchange in the $s$-channel and it gives a dominant contribution near the resonance $m_{\tilde s}^2 \simeq 4 M_{\eta_1}^2$.

We focus our analysis on the gamma rays observed at the Galactic Center \cite{mono-g1} and dwarf spheroidal galaxies \cite{mono-g2}. If we assume density distribution $\rho_{DM}$ and velocity dispersion $\bar{v}$ of DM, the observed flux of gamma rays gives a constraint on the velocity weighted thermal averaged cross section $\langle\sigma_A v\rangle_{\gamma\gamma}$ predicted by the model through Eq.~\eqref{g-spectrum}. Although DM velocity dispersion is not known well, we take it here as an example such as  $\bar{v} \sim 3 \times 10^{-4} c$ for the Galactic Center \cite{gc} and $\bar{v} \sim 3 \times 10^{-5} c$ for dwarf spheroidal galaxies, respectively. Then, we can estimate $\langle\sigma_A v\rangle_{\gamma\gamma}$ using Eq.~\eqref{g-cross} for typical model parameters. The results are shown in Table \Rnum{2}. The observation of line spectrum of gamma rays by H.E.S.S. gives constraints on $\langle\sigma_A v\rangle_{\gamma\gamma}$ at $m_{DM} = 1$ TeV such that $\langle\sigma_A v\rangle_{\gamma\gamma} < 4 \times 10^{-28}\, \mathrm{cm^3/s}$ for the Galactic Center with Einastio profile and $\langle\sigma_A v\rangle_{\gamma\gamma} < 3 \times 10^{-25}\, \mathrm{cm^3/s}$ for the dwarf spheroidal galaxies. The predicted values for the latter are found to be well below the limit. On the other hand, the former one is not far from the present bound depending on the assumed value of parameters. It may be useful to note that the behavior of $\langle\sigma_A v\rangle$ in the case (A) and (B) is found as green lines in Fig.~3. It shows that  $\langle\sigma_A v\rangle$ in the case (B) is larger than a required value by the present DM relic abundance at $x \gtrsim 10^7$. Table \Rnum{2} shows that the monochromatic gamma search could be an effective way to examine the model in such a case especially. Since the annihilation cross section is sensitive to the value of $m_{\tilde s}$ and $\kappa_2$, we may get a bound on them by using observational results of monochromatic gammas in future. If the monochromatic gammas from the Galactic Center are discovered at this energy region, this type of model with appropriate parameters could be an interesting candidate for it.

\section{Summary}
%
We extended the scotogenic neutrino mass model with a real singlet scalar to explain the origin of the right-handed neutrino mass. This extension could make the model incorporate inflation of the Universe escaping problems appearing in the Higgs inflation. The inflation could be realized naturally in the same way as the Higgs inflation. However, since the singlet scalar which is identified with inflaton is free from phenomenological constraints unlike the Higgs boson, the nonminimal coupling with Ricci scalar can take a rather small value compared with the Higgs inflation case. Moreover, there appears no unitarity violation problem in the scattering process mediated by the gravity until the inflation scale. Although reheating temperature is not high enough compared with the one required in the ordinary leptogenesis, sufficient baryon number asymmetry can be generated through leptogenesis owing to the singlet scalar.

On the other hand, the singlet scalar could change DM phenomenology substantially from the one of the original model. A DM candidate in the model is  a neutral component $\eta^0_R$ of the inert doublet, which is indispensable for the neutrino mass generation. If the singlet scalar satisfies the resonant condition with $\eta^0_R$, both the self-scattering cross section and the annihilation cross section of $\eta^0_R$ mediated by the singlet scalar could be enhanced largely through the $s$-channel singlet scalar exchange. As a result, the $\eta^0_R$ capture rate in the Sun could be enhanced through it. Since the velocity dispersion in the Galactic halo and in the Sun takes different values, it is possible that $\eta^0_R$ annihilation cross section is kept to be the required value by the DM relic abundance but only the self-scattering cross section in the Sun is enhanced. Taking account of these points, we have estimated the $\eta^0_R$ annihilation to neutrinos in the Sun and to high energy monochromatic gamma rays in the Galactic Center and dwarf spheroidal galaxies.

\newpage
\section*{Appendix A: Effective couplings of $S$ with gauge bosons}
%
The singlet scalar $S$ couples with gauge bosons $W^\pm$, $Z$ and photons $A$ through one-loop diagrams with $\eta_i$ in internal lines as shown in Fig.~1. It can be expressed as effective couplings $\sum_V \mathcal{G}^V_{\mu\nu} S^2 V_i^\mu V_i^\nu$ for $V^\mu = W^{\pm \mu},~Z^\mu$ and $A^\mu$. We present an explicit expression of $\mathcal{G}^V_{\mu \nu}$ here. We define four momenta of the final state gauge bosons as $k_1^\mu$ and $k_2^\mu$ and their polarization vectors as $\varepsilon^\mu(k_1)$ and $\varepsilon^\mu(k_2)$, respectively. The center of mass energy is $s = (k_1 + k_2)^2$ and the mass of $W^\pm$ and $Z$ is $m_V^2 = g_V^2 \langle\phi\rangle^2/2$ with $g_V = g$ and $g/c_w$, respectively. Since $M_{\eta_1} \gg \langle\phi\rangle$ and $s \gg m_V^2$ are supposed to be satisfied in the present model, this coupling can be approximately estimated as
\begin{align}
  \mathcal{G}^V_{\mu \nu}
  &\simeq \frac{\kappa_2 g_V^2}{(4 \pi)^2} \mathcal{I}\left(\frac{M_{\eta_1}^2}{s}\right)
  \left(g_{\mu \nu} - \frac{2 k_{2 \mu} k_{1 \nu}}{s}\right) \qquad
	(V = W^\pm,~Z,~A), \nonumber \\
  \mathcal{G}^V_{\mu \nu}
  &\simeq \frac{\kappa_2}{(4 \pi)^2}
	(\lambda_+ + \lambda_- + 2 \lambda_3)
  \frac{m_V^2}{s} \mathcal{J}\left(\frac{M_{\eta_1}^2}{s}\right)g_{\mu\nu} \qquad
  (V = W^\pm,~Z),
\end{align}
where $g_V = \sqrt{2} e$ for photon and $\mathcal{I}(r)$ and $\mathcal{J}(r)$ are given as
\begin{align}
  \mathcal{I}(r)
  &= 1 + r \left(\ln\frac{1 + \sqrt{1 - 4 r}}{1 - \sqrt{1 - 4 r}} + i \pi\right)^2,
  \nonumber \\
  \mathcal{J}(r)
  &= \sqrt{1 - 4 r} \left(\ln\frac{1 + \sqrt{1 - 4 r}}{1 - \sqrt{1 - 4 r}} + i \pi\right)
  - 2.
  \label{int}
\end{align}
In high energy regions where $s \gg m_V^2$ is satisfied, a dominant contribution to this coupling $\mathcal{G}_{\mu \nu}^V$ from transverse polarization $\varepsilon_T$ and longitudinal polarization $\varepsilon_L$ is approximately summarized as
\begin{align}
  &\mathcal{G}^V_{\mu \nu} \varepsilon_T^\mu(k_1, \alpha) \varepsilon_T^\nu(k_2, \beta)
  \simeq \frac{\kappa_2 g_V^2}{(4 \pi)^2}
	\mathcal{I}\left(\frac{M_{\eta_1}^2}{s}\right)
  \delta_{\alpha \beta} \qquad (V = W^\pm,~Z,~A), \nonumber \\
  &\mathcal{G}^V_{\mu \nu} \varepsilon_L^\mu(k_1) \varepsilon_L^\nu(k_2)
  \simeq \frac{\kappa_2}{2 (4 \pi)^2}(\lambda_+ + \lambda_- + 2 \lambda_3)
  \mathcal{J}\left(\frac{M_{\eta_1}^2}{s}\right) \qquad
	(V = W^\pm,~Z).
\end{align}
%

\section*{Appendix B: Capture rates of DM in the Sun}
%
We define $u$ as velocity of $\eta_1$ at infinity and $w$ as velocity of $\eta_1$ after scattering by a nucleus at a point whose distance from the center of the Sun is $r$. They satisfy $w^2 = u^2 + v^2(r)$ where $v(r)$ is escape velocity at the scattering point. If we represent the escape velocity at the center and the surface of the Sun as $v_c$ and $v_s$ respectively, the escape velocity $v(r)$ at a sphere of radius $r$ might be approximated by\footnote{They should be taken as values such that $v_c = 1354$~km/s and $v_s = 795$~km/s in this context \cite{capt}.}
\begin{align}
  \frac{v(r)^2}{v_s^2}
  = \frac{v_c^2}{v_s^2} - \frac{M(r)}{M_\odot} \left(\frac{v_c^2}{v_s^2} - 1\right),
  \label{escvel}
\end{align}
where $M(r)$ is the mass contained in the sphere of  radius $r$. We define $\hat{\phi}(r)$ as $\hat{\phi}(r) = v(r)^2/v_s^2$. If the energy transfer $\Delta E$ from $\eta_1$ to a nucleus through the scattering with the nucleus inside the Sun satisfies $\Delta E \geq M_{\eta_1} w^2/2 - M_{\eta_1} v(r)^2/2 = M_{\eta_1} u^2/2$, $\eta_1$ is captured inside the Sun. On the other hand, the kinematics of the scattering between $\eta_1$ and a nucleus $\mathcal{N}_i$ whose mass is $M_{\mathcal{N}_i}$ requires $\Delta E \leq \mu_i/\mu_{i+} E$, where $E = M_{\eta_1} w^2/2$, $\mu_i = M_{\eta_1}/M_{\mathcal{N}_i}$ and $\mu_{i\pm} = \mu_i \pm 1/2$. Taking account of this range of $\Delta E$, capture probability per a scattering $\Omega_v(w)$ for $\eta_1$ which is specified by the velocity $w$ for given $u$ and $v(r)$ can be defined as
\begin{align}
  w \Omega_v^i(w)
  = \frac{\sigma(\eta_1 \mathcal{N}_i) n_{\mathcal{N}_i} w^2}{E}
  \frac{\mu_{i+}^2}{\mu_i} \int_{E_{\mathrm{min}}^i}^{E_{\mathrm{max}}^i} F_i^2(\Delta E)
  \theta\left(\Delta E - \frac{u^2}{w^2} E\right) d(\Delta E),
  \label{cc0}
\end{align}
where nucleus number density in the Sun and $\eta_1$-$\mathcal{N}_i$ scattering cross section are expressed by $n_{\mathcal{N}_i}$ and $\sigma(\eta_1\mathcal{N}_i)$, respectively. The cross section $\sigma(\eta_1\mathcal{N}_i)$ can be calculated by applying Eq.~\eqref{elcross} to Eq.~\eqref{cN}. The form factor of the nucleus $\mathcal{N}_i$ is introduced as $F^2_i(\Delta E) = \exp(-\Delta E/E_0^i)$ where $E_0^i$ is defined as $E_0^i = 3/(2 M_{\mathcal{N}_i} R_{\mathcal{N}_i}^2)$ by using nucleus mean square radius $R_{\mathcal{N}_i} \simeq \left[0.91 (M_{\mathcal{N}_i}/\mathrm{GeV})^{1/3} + 0.3\right] \times 10^{-13}$~cm. Bounds $E_{\mathrm{max}}^i$ and $E_{\mathrm{min}}^i$ in Eq.~\eqref{cc0} are fixed as
\begin{align}
  E_{\mathrm{max}}^i = \frac{\mu_i}{\mu_{i+}^2} E, \qquad  E_{\mathrm{min}}^i = 0.
  \label{e-ela}
\end{align}
Since $M_{\eta_1} \mu_i/\mu_{i+}^2 \simeq 4 M_{\mathcal{N}_i}$ and $M_{\eta_1} \mu_{i-}^2/\mu_{i+}^2 \simeq M_{\eta_1}$ are  satisfied in the present case $ M_{\mathcal{N}_i} \ll M_{\eta_1}$, Eq.~\eqref{cc0} is reduced to
\begin{align}
  w \Omega_v^i(w)
  = \frac{\sigma(\eta_1 \mathcal{N}_i) n_{\mathcal{N}_i}}{2 M_{\mathcal{N}_i}}
  E_0^i \left[e^{-\frac{M_{\eta_1} u^2}{2 E_0^i}}
  - e^{-\frac{2 M_{\mathcal{N}_i}}{E_0^i} (u^2 + v(r)^2)}\right].
  \label{prob}
\end{align}
This suggests that only $\eta_1$ with the velocity $u \ll v(r)$ could be captured in the Sun effectively.

If we suppose that distribution of the velocity $u$ of $\eta_1$ at temperature $T$ follows the Maxwell-Boltzmann distribution function and take account of circular velocity $v_0$ of the Sun around the Galaxy Center,  the modified distribution function $f_\zeta(u)$ can be expressed as
\begin{align}
  f_\zeta(u)
  &= 4 \pi u^2 n_{\eta_1} \left(\frac{M_{\eta_1}}{2 \pi T}\right)^{3/2}
  e^{-\left(\frac{3 u^2}{2 \bar{v}^2} + \zeta^2\right)}
  \frac{\sinh2 y \zeta}{2 y \zeta} \nonumber \\
  &= \left(\frac{6}{\pi}\right)^{1/2} n_{\eta_1} \frac{y}{\bar{v} \zeta}
  \left(e^{-(y - \zeta)^2} - e^{-(y + \zeta)^2}\right),
  \label{velocity}
\end{align}
where $n_{\eta_1}$ is local number density of $\eta_1$ in the halo and a variable $y$ is defined by $y^2 = M_{\eta_1}/2 T u^2$. The velocity dispersion $\bar{v}$ of $\eta_1$ is fixed by $M_{\eta_1} \bar{v}^2/2 = 3 T/2$ and $v_0$ is taken into account through $\zeta$, which is defined by $\zeta = \sqrt{3 v_0^2/(2 \bar{v}^2)}$.

Using Eqs.~\eqref{prob} and \eqref{velocity}, the capture rate of $\eta_1$ per a unit volume in the Sun is calculated by
\begin{align}
  \frac{dC_c^i}{dV}
  = \int_{u_{\mathrm{min}}^i}^{u_{\mathrm{max}}^i} \frac{f_\zeta(u)}{u}
  w \Omega_v^i(w) du.
  \label{dcc}
\end{align}
Since $u_{\mathrm{min}}^i$ and $u_{\mathrm{max}}^i$ are fixed in the case $\mu_i \gg 1$ as
\begin{align}
  u_{\mathrm{min}}^i = 0, \qquad
  u_{\mathrm{max}}^i = \sqrt{\frac{\mu_i}{\mu_{i-}^2}} v(r),
  \label{u-el}
\end{align}
Eq.~\eqref{dcc} can be reduced to
\begin{align}
  \frac{dC_c^i}{dV}
  = \left(\frac{6}{\pi}\right)^{1/2}
  \frac{\sigma(\eta_1 \mathcal{N}_i) n_{\mathcal{N}_i} n_{\eta_1}}
  {4 M_{\mathcal{N}_i} \bar{v}} E_0^i
  \left[G(y, a_i) - G(y, b_i) e^{-(a_i - b_i) y^2}\right],
  \label{ecc}
\end{align}
where $G(y,\alpha)$ is defined by
\begin{align}
  G(y, \alpha)
  =& \left[\chi \left((-\frac{\zeta}{\sqrt{1 + \alpha}}, \frac{\zeta}{\sqrt{1 + \alpha}}
  \right)\right. \nonumber \\
  &+ \left.\chi \left(\sqrt{1 + \alpha} y^2 - \frac{\zeta}{\sqrt{1 + \alpha}},
  \sqrt{1 + \alpha} y^2 + \frac{\zeta}{\sqrt{1 + \alpha}}\right)\right]
  \frac{e^{-\frac{-\alpha \zeta^2}{1 + \alpha}}}{\sqrt{1 + \alpha}}.
\end{align}
In this formula, $a_i$ and $b_i$ are defined as $a_i = M_{\eta_1} \bar{v}^2/3 E_0^i$ and $b_i = \mu_i a_i/\mu_{i+}^2$ and a definition $\chi(z_1, z_2) = \int_{z_1}^{z_2} \exp(-z^2) dz = (\sqrt{\pi}/2) \left\{\mathrm{erf}(z_2) - \mathrm{erf}(z_1)\right\}$ is used. The total capture rate by the Sun is obtained by using the above formula as
\begin{align}
  C_c = \sum_i \int_0^{R_\odot} \frac{dC_c^i}{dV} 4 \pi r^2 dr,
  \label{fcc}
\end{align}
where $R_\odot$ is the solar radius. If we use Eq.~\eqref{escvel} in Eq.~\eqref{fcc}, we can obtain the final formula \eqref{cc} given in the text, where we use variables which are defined as
\begin{align}
  &\hat A_{i\pm} = \sqrt{1 + a_i} A_i \pm \frac{\zeta}{\sqrt{1 + a_i}}, \quad
  \check{A}_{i\pm} = \sqrt{1 + b_i} A_i \pm \frac{\zeta}{\sqrt{1 + b_i}}, \quad
   \nonumber \\
  &A_i^c = A_i(v_c), \quad A_i^s = A_i(v_s),
\end{align}
in which $A_i$ is given by $A_i^2 = (3 v(r)^2/2 \bar{v}^2) (\mu_i/\mu_{i-}^2)$.

The above calculation can also apply to self-capture rate $C_s$. In that case, we have to take account that $\eta_1$ has the $s$-channel  self-scattering process mediated by $\tilde{s}$ in addition to the one caused by a $\lambda_2$ coupling. It crucially depends on the velocity distribution around the resonance. The $\eta_1$ capture probability $\Omega_v^s(w)$ through the scattering with $\eta_1$ itself in the Sun can be expressed following  the formula for the $\eta_1$-nucleus scattering given above,
\begin{align}
  w \Omega_v^s(w)
  = \frac{n(r) \sigma_{RR} w^2}{E} \left[\tilde{E}_{\mathrm{max}}
  - \mathrm{max}\left(\tilde{E}_{\mathrm{min}},~
  \frac{1}{2} M_{\eta_1} u^2\right)\right],
  \label{scap-prob}
\end{align}
where $\sigma_{RR}$ is given in Eq.~\eqref{rr} and $n(r)$ is the number density of $\eta_1$ in the Sun. As $\tilde{E}_{\mathrm{max}}$ and $\tilde{E}_{\mathrm{min}}$ are the same as the ones in the $\eta_1$-nucleus scattering case, it can be reduced to $w \Omega_v^s(w) = \sigma_{RR} v(r)^2$. Since $\mu_+ = 1$ is satisfied in the $\eta_1$-$\eta_1$ scattering and then there is no kinematic suppression, a wider range of $u$ could substantially contribute to $dC_s/dV$ through integration for the velocity distribution of $\eta_1$. Here we note that $u_{\mathrm{max}} = v(r)$ should be satisfied to guarantee $\eta_1$ not to be ejected outside the Sun. However, $u_{\mathrm{max}}$ can be safely taken as infinity in the integration of $u$ since the escape velocity $v(r)$ in the Sun is much larger than the velocity dispersion $\bar{v}$ of $\eta_1$ in the halo. On the other hand, $u_{\mathrm{min}}$ should be fixed at $u_{\mathrm{min}} = 0$ as in the capture due to the scattering with nucleus. We obtain the formula \eqref{cs} in the text by taking account of these points and assuming that the narrow resonance condition $\Delta \gg \gamma_{\tilde{s}}$ is satisfied in the $u$ integration.

\section*{ACKNOWLEDGMENT}
%
This work is partially supported by a Grant-in-Aid for Scientific Research (C) from Japan Society for the Promotion of Science (Grant No. 18K03644).

\newpage
\bibliographystyle{unsrt}

\end{document}